\documentclass[journal]{IEEEtran}

\usepackage{graphicx,amssymb,lineno}
\usepackage{amsmath,amsfonts,amssymb}

\usepackage{algorithmic}
\usepackage[usenames]{color}
\usepackage{float}

\usepackage[linesnumbered,ruled,vlined]{algorithm2e}

\newcommand*{\TitleFont}{%
      \usefont{\encodingdefault}{\rmdefault}{}{n}%
      \fontsize{20}{20}%
      \selectfont}

\usepackage{graphicx,graphics,color,epsfig,subfigure,graphpap,rotate}
\usepackage{times, verbatim, subfigure, epsfig, graphicx, latexsym, amsmath}
\usepackage{url}
\usepackage{subfigure}
\begin{document}
%
\title{\TitleFont Exploiting Device-to-Device Communications to Enhance Spatial Reuse for Popular Content Downloading in Directional mmWave Small Cells}

\author{Yong~Niu,
        Li~Su,
        Chuhan~Gao,
        Yong~Li,~\IEEEmembership{Member,~IEEE,}
        Depeng~Jin,
        and Zhu~Han,~\IEEEmembership{Fellow,~IEEE} %
 \thanks{Copyright (c) 2015 IEEE. Personal use of this material is permitted. However, permission to use this material for any other purposes must be obtained from the IEEE by sending a request to pubs-permissions@ieee.org.}

\thanks{Y. Niu, L. Su, Chuhan Gao, Y. Li, D. Jin are with State Key Laboratory on
 Microwave and Digital Communications, Tsinghua National Laboratory for Information
 Science and Technology (TNLIST), Department of Electronic Engineering, Tsinghua
 University, Beijing 100084, China (E-mails: liyong07@tsinghua.edu.cn).} 
\thanks{
Z. Han
is with the Department of Electrical and Computer Engineering,
University of Houston, Houston, TX 77004 USA (e-mail:
zhan2@uh.edu).}
\thanks{This work was partially supported by the National Natural Science
Foundation of China (NSFC) under grant No. 61201189 and 61132002, National High Tech (863) Projects
under Grant No. 2011AA010202, Research Fund of Tsinghua University under No. 2011Z05117 and
20121087985, Tsinghua University Initiative Scientific Research Program under No. 20141081231, and Shenzhen Strategic Emerging Industry Development Special Funds under No.
CXZZ20120616141708264.}
}%

\maketitle

\vspace{-1.5cm}

\begin{abstract}

With the explosive growth of mobile demand, small cells in millimeter wave (mmWave) bands underlying the macrocell networks have attracted intense interest from both academia and industry. MmWave communications in the 60 GHz band are able to utilize the huge unlicensed bandwidth to provide multiple Gbps transmission rates. In this case, device-to-device (D2D) communications in mmWave bands should be fully exploited due to no interference with the macrocell networks and higher achievable transmission rates. In addition, due to less interference by directional transmission, multiple links including D2D links can be scheduled for concurrent transmissions (spatial reuse). With the popularity of content-based mobile applications, popular content downloading in the small cells needs to be optimized to improve network performance and enhance user experience. In this paper, we develop an efficient scheduling scheme for popular content downloading in mmWave small cells, termed PCDS (popular content downloading scheduling), where both D2D communications in close proximity and concurrent transmissions are exploited to improve transmission efficiency. In PCDS, a transmission path selection algorithm is designed to establish multi-hop transmission paths for users, aiming at better utilization of D2D communications and concurrent transmissions. After transmission path selection, a concurrent transmission scheduling algorithm is designed to maximize the spatial reuse gain. Through extensive simulations under various traffic patterns, we demonstrate PCDS achieves near-optimal performance in terms of delay and throughput, and also superior performance compared with other existing protocols, especially under heavy load. The impact of the maximum number of hops of transmission paths on its performance is also analyzed for a better understanding of the role of D2D communications.



\end{abstract}

\section{Introduction}\label{S1}

 Mobile data is growing explosively. Some industry and academic experts predict a 1000-fold demand increase by 2020 \cite{data1}. In order to meet such sharp growth, there is increasing interest in deploying small cells in higher frequency bands, such as the millimeter wave (mmWave) bands
between 30 and 300 GHz, underlying the conventional macrocell networks to significantly boost the network capacity. This deployment is usually referred to as heterogeneous cellular networks (HCNs). With huge unlicensed bandwidth (e.g., the 7 GHz spectrum between 57 GHz and 64 GHz approved by the Federal Communications Commission), small cells in the 60 GHz band have gained considerable attention from academia, industry, and standardization bodies. 60 GHz communications enable multi-gigabit data rates, and broadband applications like high-speed data transfer between devices (e.g., cameras, ipads, tablets, and notebooks), real-time streaming of both compressed and uncompressed high definition television (HDTV), wireless gigabit ethernet, and wireless gaming can be supported. Recently, rapid progress in 60 GHz
mmWave circuits, including on-chip and in-package antennas, radio frequency power amplifiers, low-noise amplifiers, voltage-controlled oscillators, mixers, and
analog-to-digital converters has paved the way for more cost-effective devices in the 60 GHz band \cite{CMOS,CMOS2,CMOS3}. Several standards have been defined for indoor wireless personal area networks (WPAN) or wireless local area networks (WLAN), for example, ECMA-387 \cite{ECMA_387} , IEEE 802.15.3c \cite{IEEE_802.15.3c}, and IEEE 802.11ad \cite{IEEE_802.11ad}.

 However, to popularize and standardize products in the 60 GHz band worldwide, several technical challenges need to be addressed. Due to high carrier frequency, 60 GHz communications suffer from high propagation loss. The free space loss in the 60 GHz band is 21.6 dB worse than 5 GHz for omnidirectional communications. Thus, directional antennas should be synthesized at both transmitter and receiver to form directional high gain beams to combat the significant propagation loss \cite{beam_training,Beamtraining2,Beamtraining3}. However, with directional transmissions, the third party nodes cannot perform carrier sensing to avoid contention with the current transmissions, which is referred to as ``deafness'' \cite{deafness}. In addition, mmWave links are also vulnerable to blockage by obstacles such as humans and furniture due to weak diffraction ability \cite{MRDMAC}. Blockage by a human penalizes the link budget by 20-30 dB.


 On the other hand, content popularity in mobile networks has been found to follow the classic Zipf's law \cite{content_popularity}. It is demonstrated that a small amount of content accounts for the majority of requests, which is very popular among the majority of users. In the 60 GHz small cells, popular content downloading is widely used in many cases, such as the Group/Broadcast communication services like police, fire, and ambulance in public safety networks, device discovery, and advertising messages broadcasting \cite{broadcasting_ref}. With directional transmissions, the wireless broadcast channel in 3G/4G networks is not feasible in the 60 GHz band \cite{content_popularity,mao}. At the same time, in the user-intensive region, there is a high probability that two user devices are located near to each other. In this case, device-to-device (D2D) communications in physical proximity can be exploited for content downloading as well as saving power and improving the spectral efficiency \cite{Yong}. D2D communications in the same carrier frequencies as today's cellular systems have significant interference to the cellular users, which limits the benefits of D2D communications. However, for D2D communications in the mmWave bands, there will be no interference with the cellular systems, and the higher achievable transmission rates in mmWave bands also further increase the benefits of D2D communications. Multi-hop D2D communications in mmWave bands can also be utilized to overcome blockage by obstacles \cite{MRDMAC,tvt_own}. In addition, in the directional communication scenario, there is less interference between links, and concurrent transmissions are enabled to unleash the potential of spatial reuse. With these fundamental differences between mmWave communications and existing systems using lower carrier frequencies (e.g., from 900 MHz to 5 GHz), new scheduling schemes or MAC protocols for popular content downloading are needed to fully reap the benefits of D2D communications in mmWave bands.

 In this paper, we propose an efficient scheduling scheme for popular content downloading in mmWave small cells, termed PCDS, where D2D communications and concurrent transmissions are fully exploited to improve transmission efficiency. In PCDS, users far from the access point (AP) receive the popular content from neighboring users in close proximity that have received the content. Meanwhile, concurrent link transmissions are exploited to significantly improve network capacity. The contributions of this paper are four-fold, which are summarized as follows.

 \begin{itemize}
 \item We design a transmission path selection algorithm to establish multi-hop transmission paths for users, aiming
 at exploiting better channel conditions (higher transmission rate) between nodes in close proximity to improve transmission efficiency. In addition, the better use of concurrent transmissions (spatial reuse) is also considered in the algorithm to enhance scheduling efficiency.

 \item After transmission path selection, we formulate the optimal multi-hop transmission scheduling problem into a mixed integer
 linear programming (MILP), i.e., to minimize the number of time slots to send the popular content from the AP to all users. Concurrent transmissions are explicitly considered under the signal to interference plus noise ratio (SINR) interference model in this formulated problem.

 \item We propose an efficient and practical concurrent transmission scheduling algorithm to solve the formulated NP-hard problem with low complexity.

 \item Extensive
 simulations under various traffic patterns are carried out to demonstrate near-optimal performance of PCDS in terms of delay and throughput, and the superior network performance of PCDS compared with other schemes. In addition, we also analyze the impact of the maximum number of hops in the transmission path selection algorithm on the performance of PCDS, which provides references for the choice of this parameter in practice.

\end{itemize}


The rest of this paper is organized as follows. The related work on directional MAC protocols for small cells in the 60 GHz band is
introduced and discussed in Section \ref{S6}. Section \ref{S2} introduces the system model, and
illustrates the procedure and problems of PCDS. Section \ref{S3} presents the proposed transmission
path selection algorithm to fully exploit D2D transmission and spatial reuse. After
transmission path selection, we formulate the problem, and propose a concurrent transmission scheduling algorithm in Section
\ref{S4}. Section \ref{S5} presents the performance evaluation of PCDS under various traffic patterns. Finally, we conclude this paper in Section \ref{S7}.

\section{Related Work}\label{S6}

To keep up with surging demand, small cells densely deployed underlying the conventional homogeneous macrocell network have been proposed to create the dual benefits of higher quality links and more spatial reuse \cite{60GHz-backhaul-2,dense cells}. However, reducing the radii of small cells in the same carrier frequencies as today's cellular systems to reap the spatial reuse benefits is fundamentally limited by interference constraints
\cite{Pico_60GHz}. By using higher frequency bands, such as the millimeter wave (mmWave) bands between 30 and 300 GHz, small cells can significantly boost the overall network capacity due to less interference with macrocells and higher achievable data rates \cite{60GHz-backhaul-2,Pico_60GHz}.

There has been some related work on directional MAC protocols for small cells in the 60 GHz band \cite{Qiao,EX_Region,Qiao_6,Qiao_15,Qiao_7}. Some work is based on TDMA \cite{ECMA_387,IEEE_802.15.3c}. Cai \emph{et al.} \cite{EX_Region} derived exclusive region (ER) conditions that concurrent transmissions always outperform TDMA. In two protocols \cite{Qiao_6, Qiao_15} based on IEEE 802.15.3c, concurrent transmissions are enabled if the multi-user interference (MUI) is below a specific threshold. Qiao \emph{et al.} \cite{Qiao} proposed a concurrent transmission scheduling algorithm for an indoor IEEE 802.15.3c WPAN, where concurrent transmissions are optimized to maximize the number of flows with the quality of service requirement of each flow satisfied. Qiao \emph{et al.} \cite{Qiao_7} also proposed a multi-hop concurrent transmission scheme to address the link outage problem and combat huge path loss to improve flow throughput. For the TDMA based protocols, unfair medium time allocation problem exists for individual users under bursty data traffic \cite{mao}.



There are also other centralized work on MAC protocols for small cells in the 60 GHz band. Gong \emph{et al.} \cite{Gong} proposed a directive CSMA/CA protocol, which solves the deafness problem by the virtual carrier sensing. However, the spatial reuse is not considered. In the multihop relay directional MAC (MRDMAC), relay paths are established to steer around obstacles \cite{MRDMAC}. However, since most transmissions are through the piconet coordinator (PNC), concurrent transmissions are also not considered in MRDMAC. Chen \emph{et al.} \cite{chen_2} proposed a spatial reuse strategy for an IEEE 802.11 ad WPAN, where two different service periods (SPs) are scheduled to overlap with each other. Since only two links are considered for concurrent transmissions, the spatial reuse is not fully exploited. Son \emph{et al.} \cite{mao} proposed a frame based directional MAC protocol (FDMAC), which amortizes the scheduling overhead over multiple concurrent transmissions in a row to achieve high efficiency. The core of FDMAC is the Greedy Coloring (GC) algorithm, which fully exploits spatial reuse and greatly improves the network throughput. FDMAC also has a good fairness performance and low complexity. Chen \emph{et al.} \cite{chenqian} proposed a directional cooperative MAC protocol (D-CoopMAC) to coordinate the uplink channel access in an IEEE 802.11ad WLAN. In D-CoopMAC, a two-hop relay path of high channel quality from the source to the destination is established to replace the direct path of poor channel quality for higher transmission efficiency. Niu \emph{et al.} \cite{tvt_own} proposed a blockage robust and efficient directional MAC protocol (BRDMAC), which overcomes the blockage problem by two-hop relaying. In BRDMAC, relay selection and spatial reuse are jointly optimized to achieve robust network connectivity and also improve network performance. Niu \emph{et al.} \cite{wcm_my} also proposed a channel transmission rate aware directional MAC protocol, RDMAC, where both the multirate capability of links and spatial reuse are exploited to improve network performance. There are two stages in RDMAC. The first stage measures the channel transmission rates of links by a heuristic algorithm, which can compute near-optimal measurement schedules with respect to the total number of measurements. The second
stage accommodates the traffic demand of links by a heuristic transmission scheduling algorithm, which can compute near-optimal transmission schedules with respect to the total transmission time. Recently, Niu \emph{et al.} \cite{JSAC_own} proposed a
joint transmission scheduling scheme for the radio access and
backhaul of small cells in 60 GHz band, termed D2DMAC, where
a path selection criterion is designed to enable D2D
transmissions for performance improvement.

There are also distributed MAC protocols for small cells in the 60 GHz band \cite{mao_15, DtDMAC}. The memory-guided directional
MAC (MDMAC) alleviates the deafness problem by incorporating a Markov state transition diagram, and employs memory to achieve approximate time division multiplexed (TDM) schedules \cite{mao_15}. The directional-to-directional MAC (DtDMAC) solves the asymmetry-in-gain problem with both senders and receivers operate in a directional-only mode \cite{DtDMAC}. DtDMAC adopts an exponential backoff procedure for
asynchronous operations, and the deafness problem is also alleviated by a Markov state transition
diagram.

To the best of our knowledge, none of the work above considers the content popularity in the problem, and we try to exploit the D2D transmission in close proximity to enhance spatial reuse for popular content downloading in the 60 GHz small cells.

\section{System Overview}\label{S2}

\subsection{System Model}\label{S2-1}

We consider a mmWave small cell of $n$ nodes, one of which is the AP, and the rest are users (UEs). The system time is partitioned into non-overlapping time slots of equal length, and the AP synchronizes the clocks of UEs and schedules the medium access of all nodes to accommodate their traffic demands. With electronically
steerable directional antennas equipped at the AP and UEs, directional transmissions are supported between any pair of nodes. In addition, a bootstrapping program is run in the system such that the AP knows the up-to-date network topology and the location information of other nodes \cite{bootstrapping,location_1}. The network topology can be obtained by the neighbor discovery schemes in \cite{bootstrapping}. Location information can be obtained based on wireless channel signatures, such as angle of arrival, time difference of arrival, or the received signal strength \cite{location_1}. We also assume all nodes are half-duplex, and each node has at most one connection with one neighbor simultaneously.

For small cells in the 60 GHz band, non-line-of-sight (NLOS) transmissions suffer from higher attenuation than line-of-sight
(LOS) transmissions \cite{NLOS,NLOS2,NLOS3}. In Ref. \cite{NLOS}, the path loss exponent in the LOS hall is 2.17, while the path loss exponent in the NLOS hall is 3.01. If the distance between the transmitter and the receiver is 10 m, the gap in the path loss is about 10 dB for the LOS hall and the NLOS hall. When operating in a power-limited regime, a 10 dB power loss requires a 10-fold reduction in transmission rate to maintain the same reliability. On the other hand, restricting to the LOS path can maximize the power efficiency since the LOS path is strongest. In addition, NLOS transmissions in the 60 GHz band also suffer from a shortage of multipath, and it is reasonably accurate to calculate link budget for a directional LOS link based on the simple additive
white Gaussian noise channel model \cite{MRDMAC}. Thus, we only consider the Line-of-Sight (LOS) transmissions to achieve high transmission rates and maximize power efficiency \cite{MRDMAC}. In \cite{MRDMAC}, it has been demonstrated that relaying using LOS links is able to obtain robust network connectivity and high throughput, and multiple schemes have been proposed to overcome blockage by multi-hop relaying \cite{tvt_own, multihop_my}.

We denote the directional link from node $i$ to $j$ by $(i,j)$, and assume the beamforming process between node $i$ and $j$ has been completed. According to the path loss model, the received power at node $j$ for $(i,j)$, ${{P^r_{ij}}}$ (mW) can be estimated by
\begin{equation}
{P^r_{ij}} = {k_0}{G_t}(i,j){G_r}(i,j){l_{ij}^{ - \tau }}{P_t},
\end{equation}
where ${{P_t}}$ (mW) denotes the transmission power, $k_0$ is a constant coefficient and proportional to ${(\frac{\lambda }{{4\pi }})^2}$ ($\lambda $ denotes the wavelength), ${G_t}(i,j)$ denotes the transmit antenna gain of node $i$ in the direction from node $i$ to $j$, ${G_r}(i,j)$ denotes the receive antenna gain of node $j$ in the direction from node $i$ to $j$, ${{l_{ij}}}$ (m) denotes the distance between transmitter $i$ and receiver $j$, and $\tau $ denotes the path loss exponent \cite{Qiao}.


Due to directional transmissions, there is less interference between links, and concurrent transmissions (spatial reuse) are enabled to improve network capacity. However, due to the limited communication range, the interference between links cannot be neglected. In this paper, we adopt the interference model in \cite{Qiao}. Then for link $(u,v)$ and $(i,j)$, the received interference power at node $j$ from node $u$ can be calculated as
\begin{equation}
I_{uvij} =\rho k_0{G_t}(u,j){G_r}(u,j){l_{uj}}^{ - \tau }{P_t},
\end{equation}
where $\rho$ denotes the MUI factor related to the cross correlation of signals from different links \cite{Qiao}. If we denote the set of links that transmit concurrently with link $(i,j)$ by $\mathbb{C}_{ij}$, then the interference power ${{I_{ij}}}$ can be calculated as
\begin{equation}
{I_{ij}} = \sum\limits_{(u,v) \in {\mathbb{C}_{ij}}} {{I_{uvij}}}.
\end{equation}
Since each node has at most one connection with one neighbor, adjacent links cannot be
scheduled concurrently \cite{mao}. Thus, $\mathbb{C}_{ij}$ should not include links that are adjacent to link $(i,j)$.
The received SINR at receiver $j$ can be calculated as

\begin{equation}
{\Gamma_{ij}} = \frac{{{k_0}{G_t}(i,j){G_r}(i,j){l_{ij}}^{ - \tau }P_t}}{{{N_0}W + \rho \sum\limits_{(u,v) \in {\mathbb{C}_{ij}}} {k_0{G_t}(u,j){G_r}(u,j){l_{uj}}^{ - \tau }{P_t}} }},
\end{equation}
where $W$ (Hz) is the bandwidth, and ${{N_0}}$ (mW/Hz) is the one-sided
power spectra density of white Gaussian noise \cite{Qiao}.

For each link $(i,j)$, we denote the minimum SINR to
support its transmission rate ${c_{ij}}$ by $\gamma ({c_{ij}})$. Therefore, link $(i,j)$'s SINR
${\Gamma_{ij}}$ should be greater than or equal to $\gamma({c_{ij}})$ to support its
concurrent transmissions with other links.


Due to the difference in distance between the transmitter and receiver, the channel transmission rates of different links vary significantly. We denote the $n \times n$ channel transmission rate matrix by ${\bf{C}}$, whose $(i,j)$ element is denoted by ${{c_{ij}}}$. ${{c_{ij}}}$ indicates the channel transmission rate of link $(i,j)$, and numerically it is equal to the number of packets that link $(i,j)$ can transmit in one time slot. We assume that a channel transmission rate measurement procedure executes in the system to update the channel transmission rate matrix \cite{wcm_my}. This process is summarized as follows: the transmitter of each link firstly transmits measurement packets to the receiver. Then with the measured signal to noise ratio (SNR), the receiver obtains the achievable transmission rate and appropriate modulation and coding scheme (MCS) according to the SNR and MCS correspondence table. Lastly, the receiver will transmit an acknowledgement packet to inform the sender of the transmission rate and MCS. Under low user mobility, the
procedure is executed periodically, and the measurement results will be reported to the AP.

\begin{figure*}[htbp]
\begin{minipage}[t]{0.5\linewidth}
\centering
\includegraphics[width=8.8cm]{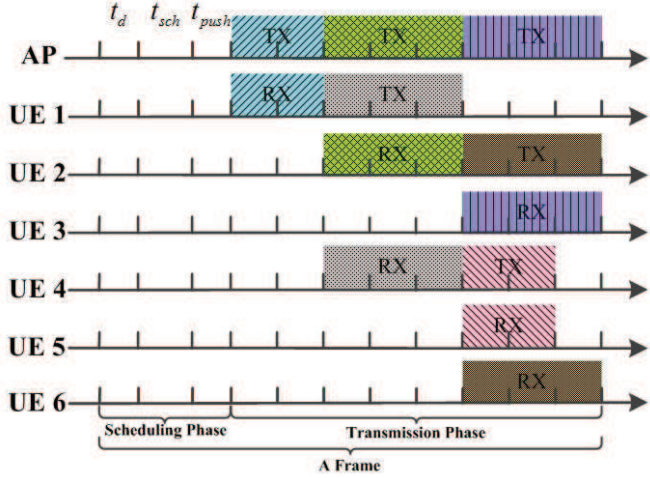}
\centerline{\small (a) Time-line illustration of PCDS operation}
\end{minipage}%
\begin{minipage}[t]{0.5\linewidth}
\centering
\includegraphics[width=7.0cm]{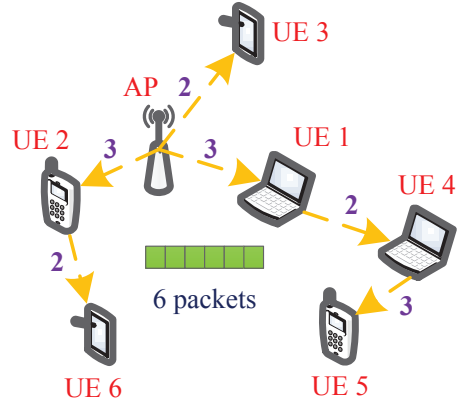}
\centerline{\small (b) The network topology}
\end{minipage}
\caption{An example of PCDS operation in a small cell of six UEs.}
\label{fig:BCTS operation} 
\vspace*{-3mm}
\end{figure*}

In this paper, we consider the popular content downloading traffic from the AP to all users. Traditionally, the traffic is distributed to each UE from the AP one by one in sequence. In PCDS, both D2D transmissions in close proximity and concurrent transmissions of links are exploited to improve transmission efficiency significantly. In Fig. \ref{fig:BCTS operation} (a), we give a time-line illustration of PCDS, where there are six UEs and an AP. In PCDS, time is divided into a sequence of non-overlapping frames \cite{mao}, and there are two phases in each frame, the scheduling phase and transmission phase. In the scheduling phase, the AP obtains the packets to be downloaded by users from the network layer, which takes time ${t_{d}}$; then the AP computes the content downloading source for each UE, and calculates a schedule to distribute these packets to all UEs, which takes time ${t_{sch}}$; lastly the AP pushes the schedule and selected downloading sources to the UEs in sequence, which takes time ${t_{push}}$. In the transmission phase, all nodes start transmissions following the schedule until the packets are distributed to all UEs. There are multiple pairings in the transmission phase, and in each pairing, multiple links are activated simultaneously for concurrent transmissions. We show an example of PCDS operation in a small cell of six UEs in Fig. \ref{fig:BCTS operation}, and the more detailed explanation of this example can be found in Section \ref{S2-2}.

%
%
%

\subsection{Problem Overview}\label{S2-2}

To improve transmission efficiency, appropriate transmission paths from the AP to UEs need to be selected to fully exploit the potential of D2D communications, and also the benefits of concurrent transmissions. After transmission paths are selected, efficient concurrent transmission scheduling algorithm should be designed to unleash the potential of spatial reuse.

Now, we consider a small cell of an AP and six UEs, whose topology is plotted in Fig. \ref{fig:BCTS operation} (b). We denote the content downloading traffic by $d$, and there are $d=6$ packets to be distributed to all UEs. The channel transmission rate matrix ${\bf{C}}$ is

\begin{equation}
{\bf{C}}=\left(
  \begin{array}{ccccccc}
    0 & 1 & 1 & 2 & 2 & 1 & 3 \\
    1 & 0 & 1 & 1 & 1 & 2 & 3 \\
    1 & 1 & 0 & 1 & 1 & 1 & 2 \\
    2 & 1 & 1 & 0 & 3 & 1 & 1 \\
    2 & 1 & 1 & 3 & 0 & 1 & 1 \\
    1 & 2 & 1 & 1 & 1 & 0 & 1 \\
    3 & 3 & 2 & 1 & 1 & 1 & 0 \\
  \end{array}
\right), \label{capacity_matrix}
\end{equation}
where the first six rows/columns are UEs, and the AP is the last row/column. We can observe that the (7,1) element of ${\bf{C}}$ is 3, which means that the link from AP to UE 1 is able to transmit 3 packets in one time slot. If we select three transmission paths, ${\rm{AP}} \to {\rm{UE1}} \to {\rm{UE4}} \to {\rm{UE5}}$, ${\rm{AP}} \to {\rm{UE2}} \to {\rm{UE6}}$, and ${\rm{AP}} \to {\rm{UE3}} $ as in Fig. \ref{fig:BCTS operation} (b), then we can obtain a schedule to complete traffic downloading of all UEs as shown in Fig. \ref{fig:BCTS operation} (a). This schedule has three pairings, and in the first pairing, the AP transmits the packets to UE 1 for two time slots. In the second pairing, the AP transmits to UE 2, and UE1 transmits the packets to UE 4 for three time slots. In the third pairing, link ${\rm{AP}} \to {\rm{UE3}}$, ${\rm{UE2}} \to {\rm{UE6}}$, and ${\rm{UE4}} \to {\rm{UE5}}$ are activated to distribute the packets to UE 3, UE 5, and UE 6. As we can see, the schedule completes traffic downloading in 8 time slots. However, if these packets are distributed from the AP to UEs one by one without exploiting the D2D communications and spatial reuse, we will need at least 25 time slots to distribute these packets to all UEs. Since selection of transmission paths has a big impact on the efficiency of spatial reuse, we should optimize the transmission paths for higher transmission efficiency. Then efficient transmission scheduling should also be investigated to fully explore the potential of spatial reuse.

\section{Transmission Path Selection}\label{S3}

In this section, we propose a heuristic transmission path selection algorithm to establish multi-hop transmission paths for better use of D2D communications and spatial reuse to improve transmission efficiency.



In the algorithm, the AP and UEs that have received the content downloading packets are scheduled to be the downloading sources of the nearest UEs to fully exploit the advantages of D2D communications. To reduce the number of adjacent links for better usage of spatial reuse in transmission scheduling, each UE is allowed to be the downloading source of one nearby UE once. The AP is allowed to be the downloading sources of multiple UEs. We denote the set of UEs in the small cell by $\mathbb{U}$. We also denote the set of UEs whose downloading sources have been selected by $\mathbb{U}_b$, and the set of UEs whose downloading sources are not selected is denoted by $\mathbb{U}_c$. When $|\mathbb{U}_b|<|\mathbb{U}_c|$, i.e., there are more UEs that are not selected than those UEs that selected, the algorithm reaches out from the nodes in $\mathbb{U}_b$ to find the UEs in $\mathbb{U}_c$ with the largest transmission rates to establish downloading links. When $|\mathbb{U}_b| \ge |\mathbb{U}_c|$, the algorithm reached out from the nodes in $\mathbb{U}_c$ to find the UEs in $\mathbb{U}_b$ with the largest transmission rates to establish downloading links.


For each UE $u \in \mathbb{U}$, we define a binary variable $b_u$ to indicate whether the downloading source of UE $u$ has been selected. If so, $b_u$ is equal to 1; otherwise, $b_u$ is equal to 0. The set of UEs whose downloading sources are selected currently is denoted by $\mathbb{U}_t$. We denote the set of the selected transmission paths by $\mathbb{P}_b$. We also denote the AP in the small cell by $\alpha$.
For each UE $u$, we define a binary variable $r_u$ to indicate whether UE $u$ has been the downloading source for other UEs. If UE $u$ has been the downloading source for other UEs, $r_u=1$; otherwise, $r_u=0$. For the AP, $r_{\alpha}$ is set to 0. For each path $p \in \mathbb{P}_b$, its number of hops is denoted by $H_p$, and its last node is denoted by $l_p$. We also define a parameter, $H_{max}$, to denote the maximum possible number of hops for each path in $\mathbb{P}_b$.


The pseudo-code of the transmission path selection algorithm is presented in Algorithm \ref{alg:path selection}. The algorithm iteratively schedules UEs into the transmission paths until all UEs are scheduled, as in line 3. When $|\mathbb{U}_b|<|\mathbb{U}_c|$, the algorithm extends the transmission paths in $\mathbb{P}_b$ by searching the neighbors of nodes in $\mathbb{U}_b$ with the largest transmission rates, as in lines 5--12. Lines 6--7 obtain the UE with the largest transmission rate from the AP to the UE, and generate a new path from the AP to the selected UE in $\mathbb{P}_b$. Lines 8--12 extend the paths in $\mathbb{P}_b$ to schedule the UEs in $\mathbb{U}_c$ into $\mathbb{P}_b$. All paths in $\mathbb{P}_b$ are restricted to have at most $H_{max}$ hops, as in line 10 and 16.
In line 10, the condition of $r_u=0$ is required since each UE is allowed to be the downloading source of one neighboring UE once. In line 12, the transmission paths in $\mathbb{P}_b$ are extended to the UE with the largest transmission rate from the last node to the node in $\mathbb{U}_c$, and $r_u$ is set to 1 since UE $u$ is the downloading source of the selected UE. When $|\mathbb{U}_b| \ge |\mathbb{U}_c|$, the algorithm extends the transmission paths in $\mathbb{P}_b$ by searching the neighbors of nodes in $\mathbb{U}_c$ with the largest transmission rates, as in lines 14--25. In lines 14--18, the set of possible downloading sources for UEs in $\mathbb{U}_c$ is denoted by $\mathbb{U}_r$, which is obtained by checking the number of hops of paths in $\mathbb{P}_b$ and whether the UE has been the downloading source for another UE, as in line 16. In lines 19--25, the downloading source for each UE in $\mathbb{U}_c$ is selected by obtaining the node in $\mathbb{U}_r$ with the largest transmission rate to the UE in $\mathbb{U}_c$. In line 26, $\mathbb{U}_t$ (the set of UEs that have been scheduled into $\mathbb{P}_b$) is added to $\mathbb{U}_b$ (the set of UEs whose downloading sources have been selected), and removed from $\mathbb{U}_c$.

\begin{algorithm}[bp!]
 \DontPrintSemicolon
 \caption{Transmission Path Selection.}\label{alg:path selection}
  \textbf{Input:} The set of UEs in the small cell $\mathbb{U}$; \\\hspace{0.95cm} Channel transmission rate matrix ${\bf{C}}$;\\
  \textbf{Initialization:} $\mathbb{U}_b=\emptyset $; $\mathbb{U}_c=\mathbb{U}$; $\mathbb{U}_t=\emptyset $; $\mathbb{P}_b=\emptyset $; $r_{\alpha}=0$; \\\hspace{2.1cm} $b_u=0$ and $r_u=0$ for each $u \in \mathbb{U}$;\\
  \While{$|\mathbb{U}_c|>0$}
  {
    $\mathbb{U}_t=\emptyset $;\\
    \If{$|\mathbb{U}_b|<|\mathbb{U}_c|$}
    {
    Obtain UE $u \in \mathbb{U}_c$ with the largest $c_{\alpha u}$; \\
    $\mathbb{P}_b$=$\mathbb{P}_b  \cup \{\alpha \to u \} $; $b_u=1$;
    $\mathbb{U}_t$=$\mathbb{U}_t  \cup \{u \}$;\\
    \For{{\rm{each}} $u\in \mathbb{U}_b$}
    {
    Obtain the path $p\in \mathbb{P}_b$ with $l_p=u$;\\
     \If{$r_u=0$ {\rm{and}} $H_p<H_{max}$}
     {
     Obtain UE $v \in \mathbb{U}_c$ with the largest $c_{uv}$ and $b_v=0$; \\
     Extending $p \in \mathbb{P}_b $ to $v$;
     $b_v=1$; $r_u=1$; $\mathbb{U}_t$=$\mathbb{U}_t  \cup \{v \}$;\\

     }
    }
    }
    \Else {
    $\mathbb{U}_r=\emptyset $;\\
    \For{{\rm{each}} $p\in \mathbb{P}_b$}
    {
    \If{(${H_p} < {H_{\max }}\;{\rm{and}}\;{r_{{l_p}}} = 0$)}
    {
        $\mathbb{U}_r=\mathbb{U}_r \cup \{l_p\} $;\\
    }
    }
    $\mathbb{U}_r=\mathbb{U}_r \cup \{ \alpha\}$;\\
    \For{{\rm{each}} $u\in \mathbb{U}_c$}
    {
    Obtain UE $v\in \mathbb{U}_r$ with the largest $c_{vu}$ and $r_v=0$;

     \If{$v$ {\rm{is}} $\alpha$}
     {
        $\mathbb{P}_b$=$\mathbb{P}_b  \cup \{\alpha \to u \} $;
        $b_u=1$; $\mathbb{U}_t$=$\mathbb{U}_t \cup \{u \}$;\\
     }
     \Else
     {
        Obtain the path $p\in \mathbb{P}_b$ with $l_p=v$;\\
        Extending $p \in \mathbb{P}_b $ to $u$;
        $b_u=1$; $r_v=1$; $\mathbb{U}_t$=$\mathbb{U}_t  \cup \{u \}$;\\
     }
    }

    }

    $\mathbb{U}_b=\mathbb{U}_b \cup \mathbb{U}_t$; $\mathbb{U}_c=\mathbb{U}_c - \mathbb{U}_t$; \\
  }

$\mathbf{Return}\ \mathbb{P}_b$.
\end{algorithm}


Applying the transmission path selection algorithm with $H_{max}=3$ to the example in Section \ref{S2-2}, we obtain the transmission paths as path ${\rm{AP}} \to {\rm{UE1}} \to {\rm{UE4}} \to {\rm{UE5}}$, path ${\rm{AP}} \to {\rm{UE2}} \to {\rm{UE6}}$, and path ${\rm{AP}} \to {\rm{UE3}} $, which have been illustrated in Fig. \ref{fig:BCTS operation} (b). The computational complexity of Algorithm \ref{alg:path selection} is $\mathcal{O}(|\mathbb{U}|^2)$.



\section{Concurrent Transmission Scheduling}\label{S4}

After transmission path selection, we schedule the downloading links for each UE in the transmission phase. The content downloading for all UEs needs to be completed with the minimum number of time slots to maximize transmission efficiency. In this section, we first formulate the optimal multi-hop transmission scheduling problem into a mixed integer program based on the problem formulation in FDMAC (the frame-based scheduling directional MAC
protocol) \cite{mao}, and then propose a practical concurrent transmission scheduling algorithm to fully exploit spatial reuse for maximizing transmission efficiency.

\subsection{Problem Formulation} \label{S4-1}


We denote the transmission schedule in the transmission phase by $\textbf{S}$, which has $K$ pairings. For each UE $u$, we denote its downloading source by $s_u$, and $s_u$ may be the AP or other UEs. The traffic demand to be distributed to all UEs is denoted by $d$. For each pairing, we define an $n \times (n - 1)$ matrix $\textbf{A}^k$ to indicate the links scheduled to communicate in the $k$th pairing. Since we focus on popular content downloading in this paper, where the traffic is from the AP to UEs, the links from users to the AP are not considered in matrix $\textbf{A}^k$. The rows of $\textbf{A}^k$ indicate the downloading sources of UEs, including all UEs and the AP, while the columns of $\textbf{A}^k$ indicate all UEs. The $(i,j)$ element of $\textbf{A}^k$, $a_{ij}^k$ indicates whether link $(i,j)$ is scheduled for transmission in the $k$th pairing. If link $(i,j)$ is scheduled for transmission in the $k$th pairing, $a_{ij}^k=1$; otherwise, $a_{ij}^k=0$. We denote the number of time slots for the $k$th pairing by $\delta ^k$ \cite{mao}. To maximize transmission efficiency, the transmission schedule should complete the traffic downloading for all UEs with a minimum number of time slots. Therefore, the objective function to be minimized is $\sum\limits_{k = 1}^K {{\delta ^k}}$ \cite{mao}. Now, we analyze the system constraints of this problem.

First, all UEs can download the traffic once in the schedule, which can be expressed as follows.

\begin{equation}
\sum\limits_{k = 1}^K {a_{{s_u}u}^k = 1}, \ \forall \;u. \label{CONS1}
\end{equation}

Second, the schedule should complete the traffic downloading for all UEs, which can be expressed as follows.

\begin{equation}
\sum\limits_{k = 1}^K {({\delta ^k} \cdot a_{{s_u}u}^k \cdot {c_{{s_u}u}}} ) \ge d,\ \forall\;u. \label{CONS2}
\end{equation}

Third, since UE $u$ obtains the common packets after UE $s_u$ received the common packets, the downloading of UE $s_u$ should be scheduled ahead of the downloading for UE $u$, which can be formulated as follows. This constraint represents a group of constraints since $\widetilde K$ varies from 1 to $K$,

\begin{equation}
\sum\limits_{k = 1}^{\widetilde K} {a_{{s_{{s_u}}}{s_u}}^k \ge } \sum\limits_{k = 1}^{\widetilde K} {a_{{s_u}u}^k},\ \forall\; u,\; \widetilde K = 1\thicksim K. \label{CONS3}
\end{equation}

Fourth, due to the half-duplex assumption, adjacent links cannot be scheduled for concurrent transmissions. Thus, the links that share common nodes cannot be scheduled in the same pairing, which can be expressed as follows.

\begin{equation}
a_{{s_u}u}^k + a_{{s_v}v}^k \le 1,\ {\rm{if}}\;({s_u},u)\;{\rm{and}}\;({s_v},v)\;{\rm{are}}\;{\rm{adjacent.}} \label{CONS4}
\end{equation}

Finally, to enable concurrent transmissions, the SINR of each link in the same pairing should be able to support its transmission
rate, which can be formulated as follows.

\begin{equation}
\begin{array}{l}
\frac{{{k_0}{G_t}(s_u,u){G_r}(s_u,u){l_{{s_u}u}}^{ - \tau }{P_t}a_{{s_u}u}^k}}{{W{N_0} + \rho \sum\limits_{{s_v}} {\sum\limits_v {{k_0}{G_t}(s_v,u){G_r}(s_v,u){l_{{s_v}u}}^{ - \tau }{P_t}a_{{s_v}v}^k} } }} \ge \gamma ({c_{{s_u}u}}) \\\times a_{{s_u}u}^k,\ \forall \;u,k.
\end{array} \label{CONS5}
\end{equation}
If link $(s_u,u)$ is not scheduled in the $k$th pairing, $a_{{s_u}u}^k$ is 0, and this constraint does not apply. Otherwise, the SINR of link $(s_u,u)$ should be greater than or equal to $\gamma({c_{{s_u}u}})$.

Therefore, the problem of optimal transmission scheduling (P1) can be formulated as follows.

\begin{equation}\hspace{-2.8cm}
({\rm{P1}})\ \ \min \sum\limits_{k = 1}^K
{{\delta ^k}} ,\label{OBJ}
\end{equation}

\hspace{2.35cm}s. t.
\hspace{0.2cm}Constraints (\ref{CONS1})--(\ref{CONS5}).

\subsection{Problem Reformulation}\label{S4-2}

Since constraints (\ref{CONS2}) and (\ref{CONS5}) are nonlinear, problem P1 is a mixed integer
nonlinear programming (MINLP), which is generally NP-hard. By the Reformulation-Linearization Technique (RLT) \cite{mao,RLT}, we linearize constraints (\ref{CONS2}) and (\ref{CONS5}). The RLT procedure is used to produce tight linear programming relaxations for an
underlying nonlinear and non-convex polynomial programming problem.
In the RLT procedure, for each nonlinear term, a
variable substitution is applied to linearize the objective function
and the constraints. In addition, nonlinear implied constraints for each
substitution variable are generated by taking the products of
bounding terms of the decision variables, up to a suitable order \cite{mao}.

For the second order term in constraint (\ref{CONS2}), we define a substitution variable $\xi _{{s_u}u}^k = {\delta ^k} \cdot a_{{s_u}u}^k$. $\delta ^k$ is bounded as $0\le{\delta ^k}\le{\overline T }$, where ${\overline T } = \max \{ \left\lceil
{\frac{{{d_{}}}}{{{c_{s_{u}u}}}}} \right\rceil, \forall \;u\} $ \cite{mao}. With $0\le{a_{{s_u}u}^k} \le 1$, we can obtain the \emph{RLT bound-factor product constraints} for $\xi _{{s_u}u}^k$ as

\begin{equation}
\left\{ {\begin{aligned}&{
{\xi _{{s_u}u}^k \ge 0}}\\
&{{\delta ^k} - \xi _{{s_u}u}^k \ge 0}\\
&{\overline T  \cdot a_{{s_u}u}^k - \xi _{{s_u}u}^k \ge 0}\\
&{ \overline T- {\delta ^k} - \overline T  \cdot a_{{s_u}u}^k + \xi _{{s_u}u}^k \ge 0 }
\end{aligned}\;\forall \;u,k}. \right.\\
\label{RLT CONS3}
\end{equation}


For constraint (\ref{CONS5}), we first convert it to

\begin{equation}
\begin{aligned}&
({k_0}{G_t}(s_u,u){G_r}(s_u,u){l_{{s_u}u}}^{ - \tau }{P_t} - \gamma({c_{{s_u}u}})W{N_0}) \times
a_{{s_u}u}^k \\&\ge \gamma({c_{{s_u}u}})\rho  \sum\limits_{{s_v}}
{\sum\limits_v
{{k_0}{G_t}(s_v,u){G_r}(s_v,u){l_{{s_v}u}}^{ - \tau }{P_t}a_{{s_u}u}^ka_{{s_v}v}^k} },\\&\ \forall \;u,k.
\end{aligned}
\end{equation}

For the second order term $a_{{s_u}u}^ka_{{s_v}v}^k$, we define $\omega_{{s_u}u{s_v}v}^k=a_{{s_u}u}^ka_{{s_v}v}^k$ as the substitution
variable. Since $0 \le a_{{s_u}u}^k \le 1$ and $0 \le a_{{s_v}v}^k \le 1$, the \emph{RLT bound-factor product constraints} for $\omega_{{s_u}u{s_v}v}^k$ are

\begin{equation}\hspace{0cm}
\left\{ {\begin{aligned}&{
{\omega_{{s_u}u{s_v}v}^k \ge 0}}\\
&{ a_{{s_u}u}^k - \omega_{{s_u}u{s_v}v}^k \ge 0}\\
&{a_{{s_v}v}^k - \omega_{{s_u}u{s_v}v}^k \ge 0}\\
&{1-a_{{s_u}u}^k - a_{{s_v}v}^k + \omega_{{s_u}u{s_v}v}^k \ge 0 }
\end{aligned}\;\forall \;u,v,k.}\right.\\\label{RLT bound-facotr constraints_2}
\end{equation}

By substituting $\xi _{{s_u}u}^k$ and $\omega_{{s_u}u{s_v}v}^k$ into constraints (\ref{CONS2}) and (\ref{CONS5}), problem P1 can be reformulated into a mixed integer linear programming (MILP) as


\begin{equation}\hspace{-7cm}
 \min \sum\limits_{k = 1}^K
{{\delta ^k}} \label{OBJ_RF}
\end{equation}
\hspace{0.14cm}s. t.

\begin{equation}\hspace{-4.50cm}
\sum\limits_{k = 1}^K {(\xi _{{s_u}u}^k \cdot {c_{{s_u}u}}} ) \ge d,\ \forall\;u; \label{CONS2-RLT}
\end{equation}

\begin{equation}\hspace{0.0cm}
\begin{aligned}&
({k_0}{G_t}(s_u,u){G_r}(s_u,u){l_{{s_u}u}}^{ - \tau }{P_t} - \gamma({c_{{s_u}u}})W{N_0}) \times
a_{{s_u}u}^k \\&\ge \gamma({c_{{s_u}u}})\rho  \sum\limits_{{s_v}}
{\sum\limits_v
{{k_0}{G_t}(s_v,u){G_r}(s_v,u){l_{{s_v}u}}^{ - \tau }{P_t}\omega_{{s_u}u{s_v}v}^k} },\\&\ \forall\;u,k;
\end{aligned}
\end{equation}
\hspace{0.2cm}Constraints (\ref{CONS1}), (\ref{CONS3}), (\ref{CONS4}), (\ref{RLT CONS3}), (\ref{RLT bound-facotr constraints_2}).\\

%

Considering the example in Section \ref{S2-2}, with the selected transmission paths by Algorithm \ref{alg:path selection}, we solve the MILP using an open-source MILP solver, YALMIP \cite{yalmip}. The optimal schedule consists of three pairings, and completes traffic downloading with eight time slots, which has been illustrated in Fig. \ref{fig:BCTS operation} (a). However, problem P1 has $\mathcal{O}((|\mathbb{U}| )^2K)$ decision variables, and $\mathcal{O}((|\mathbb{U}| )^2K)$ constraints, and using the optimization software to solve the problem takes significant computation time, which is unsuitable for mmWave systems \cite{mao}. Therefore, to implement efficient concurrent transmission scheduling in practical mmWave small cells, heuristic concurrent transmission scheduling algorithms with low computational complexity are needed, which will be constructed in the following subsection.

\subsection{Concurrent Transmission Scheduling Algorithm}\label{S4-3}

After the transmission path selection by Algorithm \ref{alg:path selection}, we propose a heuristic concurrent transmission scheduling algorithm to compute the near-optimal transmission schedules with much lower complexity than that of optimization software, borrowing the design ideas of the Greedy Coloring (GC) algorithm. Since adjacent links cannot be scheduled concurrently in the same pairing, the set of links in each pairing can be represented by a matching, and thus the maximum number of links in the same pairing is $\left\lfloor {n/2} \right\rfloor $ \cite{mao}. We denote the set of links scheduled in the $t$th pairing by $\mathbb{E}^t$, and the set of vertices of the links in $\mathbb{E}^t$ is denoted by $\mathbb{V}^t$. Thus, the problem of optimal transmission scheduling is to obtain the matching in each pairing to complete traffic downloading with a minimum of time slots. In each pairing, our algorithm first obtains transmission paths with the largest number of hops, and then the hop with the largest weight among the first unscheduled hops of these paths will be visited first \cite{mao}. To maximize spatial reuse, the algorithm iteratively allocates as many links as possible into each pairing with the concurrent transmission conditions satisfied. For the hops on the same path, the preceding hops should be scheduled first since each UE can be the downloading source for other UEs only after it has received the common packets.

We denote the set of transmission paths selected by Algorithm \ref{alg:path selection} by $\mathbb{P}_b$. For each path $p \in \mathbb{P}_b$, we denote its number of hops by $H_p$. The set of hops in $\mathbb{P}_b$ is denoted by $\mathbb{E}_b$. We denote the $h$th hop of path $p$ by $(p,h)$, and define its weight $w_{ph}$ as the number of time slots to complete traffic downloading. We denote the first unscheduled hop on path $p$ by $(p,F_p)$, where $F_p$ indicates its hop number. We also denote the transmitter of $(p,F_p)$ by $s_{pF_p}$, and the receiver by $r_{pF_p}$. In the $t$th pairing, we denote the set of paths that are not visited yet by $\mathbb{P}^t_u$.



The pseudo-code of the concurrent transmission scheduling algorithm is presented in Algorithm \ref{alg:CTS_B}. After obtaining the set of the selected transmission paths, $\mathbb{P}_b$, we obtain the set of hops in $\mathbb{P}_b$ as $\mathbb{E}_b$. The algorithm iteratively schedules the hops in $\mathbb{E}_b$ into each pairing until all hops in $\mathbb{E}_b$ are scheduled, as in line 6. In each pairing, we first visit the paths with the largest number of unscheduled hops, as in line 11. Then the first unscheduled hop with the largest weight is visited and selected as the candidate hop of this pairing, as in line 12. In line 13, the algorithm examines whether the candidate hop is adjacent to the hops already in this pairing. If the candidate hop is not adjacent to the hops already in this pairing, this candidate hop will be added to this pairing to check whether the concurrent transmission conditions of this pairing are satisfied, as in line 14 and lines 15--18. If the SINR of one link in this pairing cannot support its transmission rate, the candidate hop will be removed from this pairing, as in line 17 and 23. Otherwise, the number of time slots of this pairing is updated to accommodate the traffic demand of this candidate hop, and this hop is removed from $\mathbb{E}_b$ in line 19. The visited path is removed from $\mathbb{P}^t_u$ in line 24. If the number of links in each pairing reaches $\left\lfloor {n/2} \right\rfloor $ or there is no path unvisited, the algorithm will start scheduling for the next pairing as in line 10, and the scheduling results for this pairing will be outputted in line 25.

 \begin{algorithm}[bp!]
 \DontPrintSemicolon
 \caption{Concurrent Transmission Scheduling.}\label{alg:CTS_B}
  \textbf{Input:} The set of selected transmission paths, $\mathbb{P}_b$; \\
 \hspace{1.08cm}The set of hops in $\mathbb{P}_b$, $\mathbb{E}_b$;\\ \hspace{0.95cm} The number of hops for each path $p\in \mathbb{P}_b$, $H_p$; \\
 \hspace{0.95cm} The weight of each hop $(p,h)\in \mathbb{E}_b$, $w_{ph}$;\\
  \textbf{Initialization:} Set $F_p=1$ for each $p\in \mathbb{P}_b$; $t$=0;\\
    \While {$|\mathbb{E}_b| > 0$}
    {
        $t$=$t$+1;  \\
        Set ${\mathbb{V}^t} = \emptyset $, ${\mathbb{E}^t} = \emptyset $, and $\delta^{t}=0$; \\ Set $\mathbb{P}_u^t$ with $\mathbb{P}_u^t = \mathbb{P}_b$; \\
        \While {$|{\mathbb{P}_u^t}|>0$ {\rm{and}} $|{\mathbb{E}^t}| < \left\lfloor {n/2} \right\rfloor $}
        {
            Obtain the set of unvisited paths with the largest number of unscheduled hops, $\mathbb{P}_{mh}$; \\
            Obtain the hop $(p, F_p)$  of path $p\in{\mathbb{P}_{mh}}$ with the largest weight, $w_{pF_p}$; \\
            \If {$s_{pF_p} \notin {\mathbb{V}^t}$ {\rm{and}} $ r_{pF_p} \notin {\mathbb{V}^t}$}
            {
                ${\mathbb{E}^t} = {\mathbb{E}^t} \cup \{ (p, F_p)\}$;  ${\mathbb{V}^t} = {\mathbb{V}^t} \cup \{ s_{pF_p},r_{pF_p}\} $;\\
                \For {{\rm{each link}} $(p,h)$ {\rm{in}} ${\mathbb{E}^t}$}
                {
                    Calculate the SINR of link $(p,h)$, ${\Gamma_{ph}}$; \\
                    \If {${\Gamma_{ph}}<\gamma({c_{ph}})$}
                    {
                        Go to line 23\\
                    }
                }

                $\delta^t={\rm{max}}(\delta^t, w_{pF_p})$, $\mathbb{E}_b=\mathbb{E}_b-(p,F_p)$; \\
                \If {$F_{p}==H_{p}$}
                {
                    $\mathbb{P}_b = \mathbb{P}_b - p_{}$;\\
                }
                $F_{p}=F_{p}+1$; Go to line 24\\
                ${\mathbb{E}^t} = {\mathbb{E}^t} - \{ (p, F_p)\}$;  ${\mathbb{V}^t} = {\mathbb{V}^t} - \{ s_{pF_p},r_{pF_p}\}$;\\

            }
            $\mathbb{P}_u^t = \mathbb{P}_u^t - p_{}$;\\

        }
    Output $\mathbb{E}^t$ and ${\delta ^t}$;\\

    }

$\mathbf{Return}\ \mathbb{E}^t$ and ${\delta ^t}$ for each pairing.
\end{algorithm}

For the example in Section \ref{S2-2}, with the transmission paths selected by Algorithm \ref{alg:path selection}, Algorithm \ref{alg:CTS_B} gives the same schedule as YALMIP \cite{yalmip}. However, since the outer while loop has $|\mathbb{E}_b|$ iterations, which at most is $|\mathbb{U}|$, and the inner while loop and for loop have $|{\mathbb{E}^t}|$ iterations, which is at most $\left\lfloor {n/2} \right\rfloor $, our algorithm has the computational complexity of $\mathcal{O}(|\mathbb{U}|^3)$, which is much lower than YALMIP.

\section{Performance Evaluation}\label{S5}

In this section, we evaluate the delay and throughput of our proposed popular content downloading scheme under various traffic patterns. We compare it with two existing schemes, and also investigate the impact of the maximum number of hops $H_{max}$ on the performance of our scheme.

\subsection{Simulation Setup}\label{S5-1}

In the simulation, we consider a typical mmWave small cell of an AP and ten UEs. We assume the AP is located in the center of a square area with $10 m \times 10 m$, and the ten UEs are uniformly distributed in the area. Adopting the simulation parameters in Table II of \cite{MRDMAC}, the duration of a time slot is set to 5 $\mu s$. The packet size is set to 1000 bytes. According to the distances between nodes, we set three transmission rates, 2 Gbps, 4 Gbps, and 6 Gbps. With a transmission rate of 2 Gbps, a packet can be transmitted in a time slot \cite{mao}. The AP obtains the common packets from the upper layer or pushes the schedule to UEs in one time slot \cite{mao}. For the simulated small cell, it takes a few time slots for the AP to compute the transmission paths and transmission schedule. In the simulation, we assume nonadjacent links are able to be scheduled for concurrent transmissions.

For the downloading traffic, we adopt two modes:

\subsubsection {\textbf{Poisson Process}} The packets arrive following a Poisson process with arrival rate $\lambda $. The traffic load, denoted by
${T_p}$, is defined as
\begin{equation}
{T_p} = \frac{{\lambda  \times L \times |\mathbb{U}|}}{R}, \label{Tl_1}
\end{equation}
where $L$ denotes the size of data packets, $|\mathbb{U}|$ denotes the number of UEs, and $R$ is set to 2
Gbps.

\subsubsection {\textbf{Interrupted Poisson Process (IPP)}} The packets arrive following an interrupted Poisson
process (IPP). The parameters of the interrupted Poisson process are ${{\lambda _1}}$, ${{\lambda
_2}}$, ${{p_1}}$ and ${{p_2}}$, and the arrival intervals of an IPP obey the second-order
hyper-exponential distribution with a mean of
\begin{equation}
E(X) = \frac{{{p_1}}}{{{\lambda _1}}} + \frac{{{p_2}}}{{{\lambda _2}}}.
\end{equation}
Since the IPP can also be represented by an ON-OFF process, IPP traffic is
typical bursty traffic. The traffic load ${T_i}$ in this mode is defined as
 \begin{equation}
{T_i} = \frac{{L \times |\mathbb{U}|}}{{E(X) \times R}}.\label{Tl_2}
\end{equation}

The simulation length is ${10^5}$ time slots, and the delay threshold is set to $2.5 \times {10^4}$ time slots. The packets with delay larger than the threshold are discarded by UEs. We evaluate the system by the following three performance metrics:

1) \textbf{Average Transmission Delay:} The average traffic downloading delay from the AP to UEs, which is in units of time slots.

2) \textbf{Network Throughput:} The number of successfully transmitted packets to all UEs until end of the simulation. With the constant simulation length and fixed packet size, the total number of successfully transmitted packets is a good indication to show the throughput performance.

3) \textbf{D2D Ratio:} The fraction of packets transmitted by device-to-device links over the total number of successfully transmitted packets. This metric is used to evaluate the role of D2D communications in the traffic downloading.

In the simulation, we compare our scheme with the following two transmission schemes:

1) \emph{\textbf{SBTS}}: The serial broadcasting transmission scheme, where the packets are transmitted to UEs from the AP serially without exploiting the D2D communications.

2) \emph{\textbf{FDMAC-H}}: In FDMAC-H, the concurrent transmission scheduling algorithm borrows the design ideas of
the greedy coloring (GC) algorithm in FDMAC \cite{mao}, and performs edge coloring iteratively on the first unscheduled hops of transmission paths in a non-increasing order of weight with the conditions for concurrent transmissions satisfied. After each coloring, the set of first unscheduled hops is updated since the hops after these scheduled links become the first unscheduled hops. In FDMAC-H, the transmission path selection algorithm is the same as PCDS, and the detailed pseudo-code illustration can be found in \cite{JSAC_own}.

\subsection{Comparison with the Optimal Solution}

To show the gap between the concurrent transmission scheduling algorithm and the optimal solution of the MILP, we first compare PCDS (Section V.C.) with the optimal solution of the MILP (Section V.B.). Since obtaining the optimal solutions is time-consuming, we simulate a smaller scenario of six UEs. All other simulation parameters are as defined in Section \ref{S5-1}.

\begin{figure}[htbp]
\begin{minipage}[t]{0.5\linewidth}
\centering
\includegraphics[width=1\columnwidth,height=1.15in]{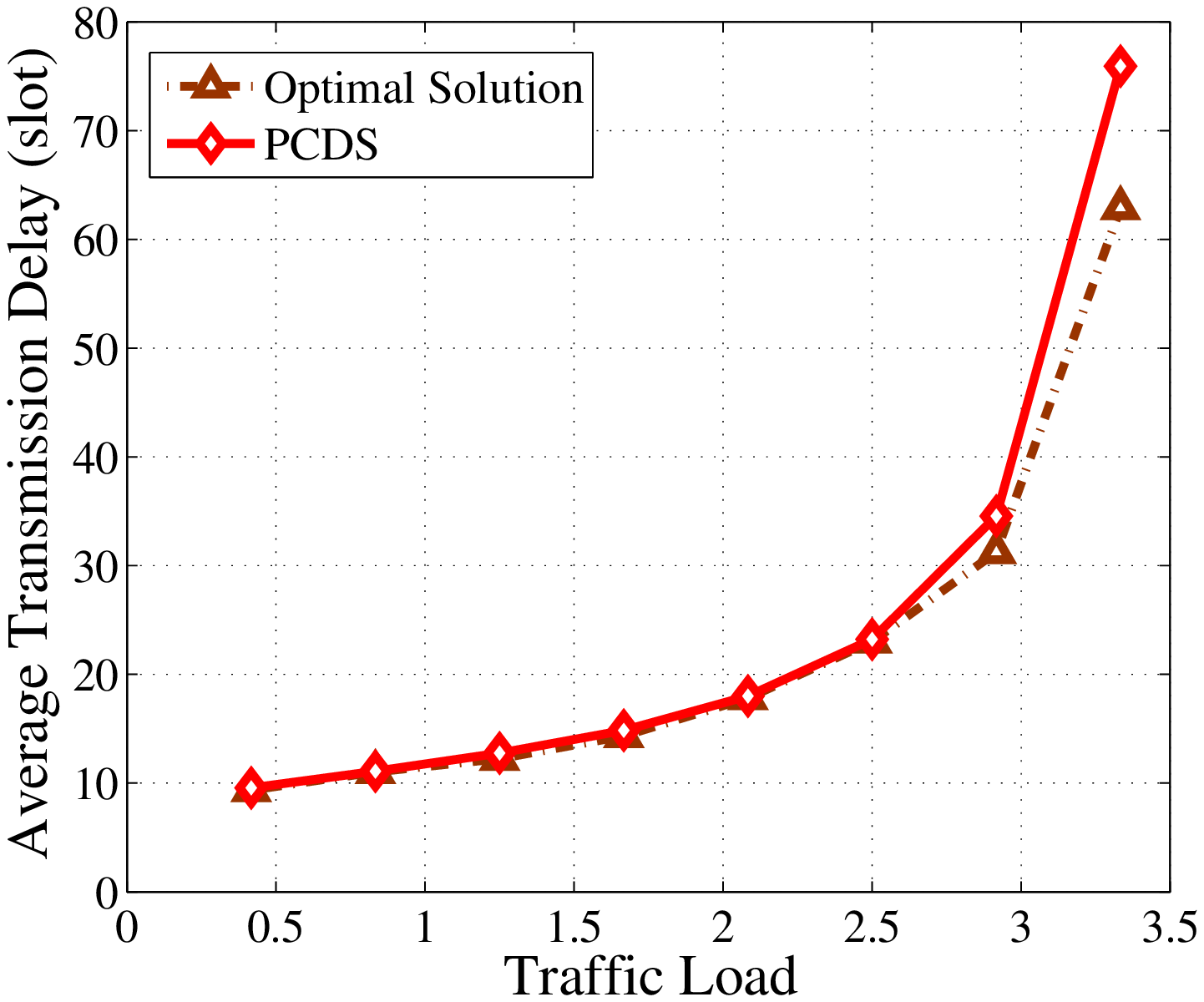}
\centerline{\small (a) Average transmission delay}
\end{minipage}%
\begin{minipage}[t]{0.5\linewidth}
\centering
\includegraphics[width=1\columnwidth,height=1.15in]{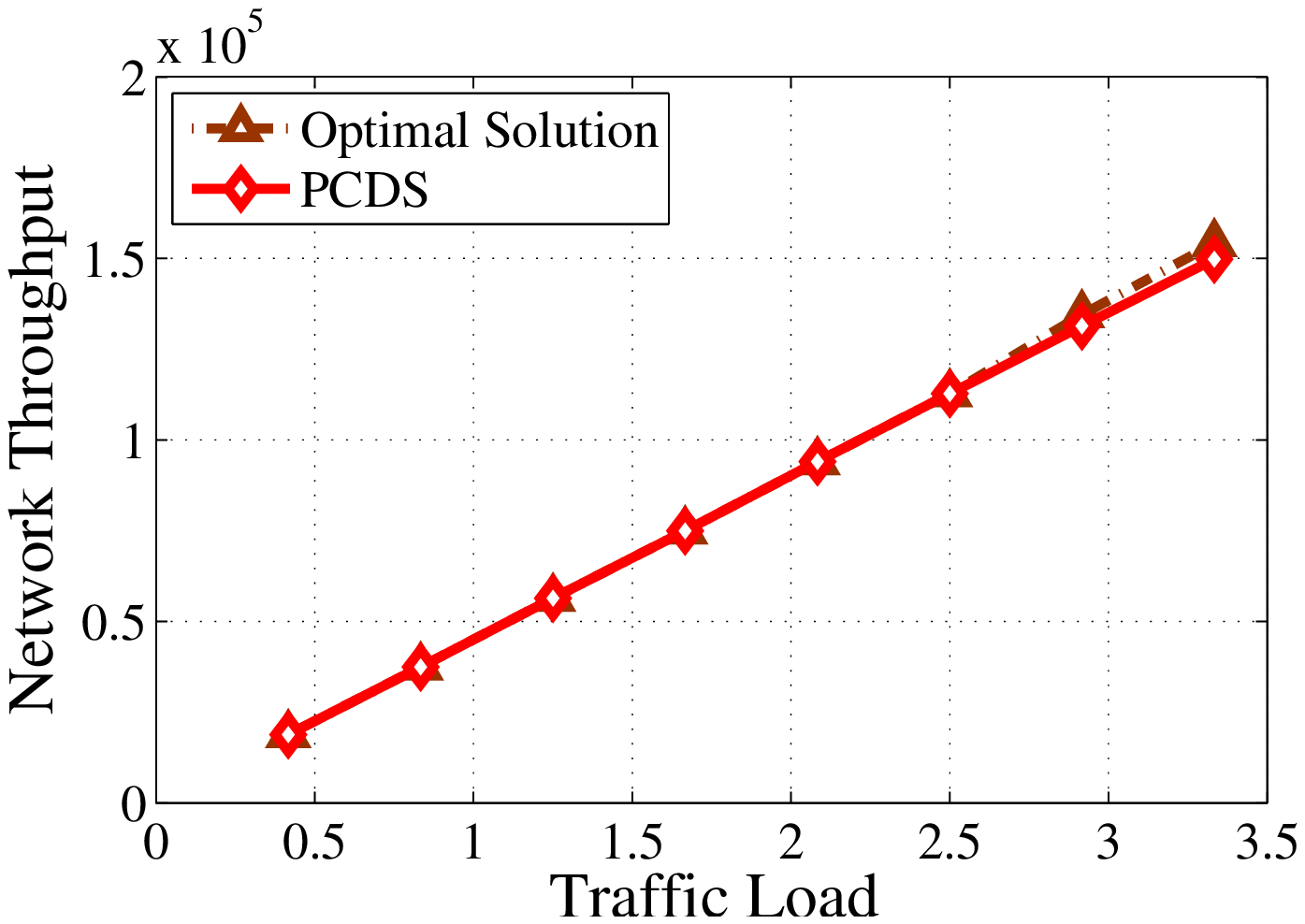}
\centerline{\small (b) Network throughput}
\end{minipage}%
\caption{Delay and throughput comparison of Optimal Solution and PCDS.}
\label{fig:comparison_opt} 
\vspace*{-3mm}
\end{figure}

We plot the delay and throughput comparison of PCDS and the optimal solution under Poisson
traffic in Fig. \ref{fig:comparison_opt}. From the results, we can observe that the gap between the delay of PCDS and the optimal solution is small under light load, and the gap increases slowly with the traffic load. The gap is only about 17.2\% at the traffic load of 3.33. In
terms of network throughput, the gap is also very small. The gap is only about 2.8\% at the traffic load of 3.33. Therefore, we have demonstrated that PCDS achieves near-optimal
performance in terms of concurrent transmission scheduling.

\begin{figure}[htbp]
\begin{minipage}[t]{1\linewidth}
\centering
\includegraphics[width=0.7\columnwidth]{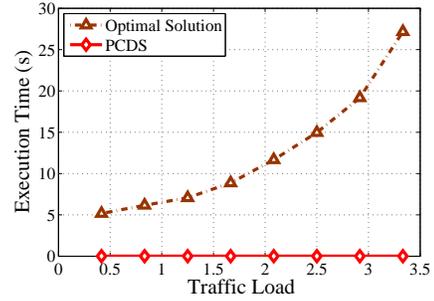}
\end{minipage}%
\caption{Execution time comparison of Optimal Solution and PCDS.}
\label{fig:time_opt} 
\vspace*{-3mm}
\end{figure}

We also plot the average execution time of PCDS and the optimal solution under Poisson traffic in
Fig. \ref{fig:time_opt}. We can observe that
the optimal solution takes much longer to execute than PCDS, and the gap increases with increasing traffic load, which indicates that PCDS has much lower computational complexity.

\subsection{Comparison with Existing Schemes} \label{S5-2}

We compare the delay and throughput of SBTS, FDMAC-H, and PCDS under different traffic patterns. The maximum number of hops of transmission paths in Algorithm \ref{alg:path selection}, $H_{max}$ is set to 4.

\subsubsection{Delay}

\begin{figure}[htbp]
\begin{minipage}[t]{0.5\linewidth}
\centering
\includegraphics[width=1\columnwidth,height=1.15in]{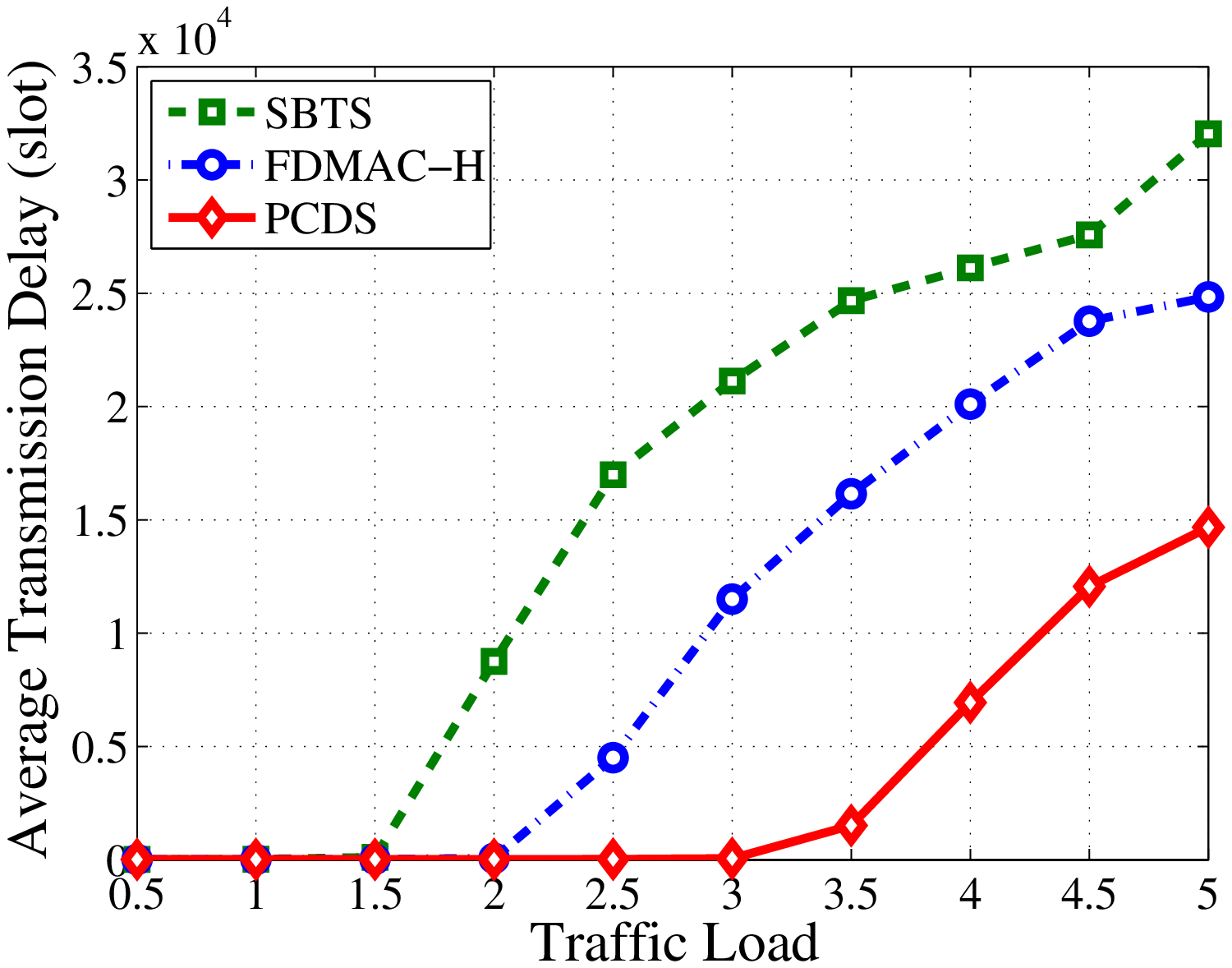}
\centerline{\small (a) Poisson traffic}
\end{minipage}%
\begin{minipage}[t]{0.5\linewidth}
\centering
\includegraphics[width=1\columnwidth,height=1.15in]{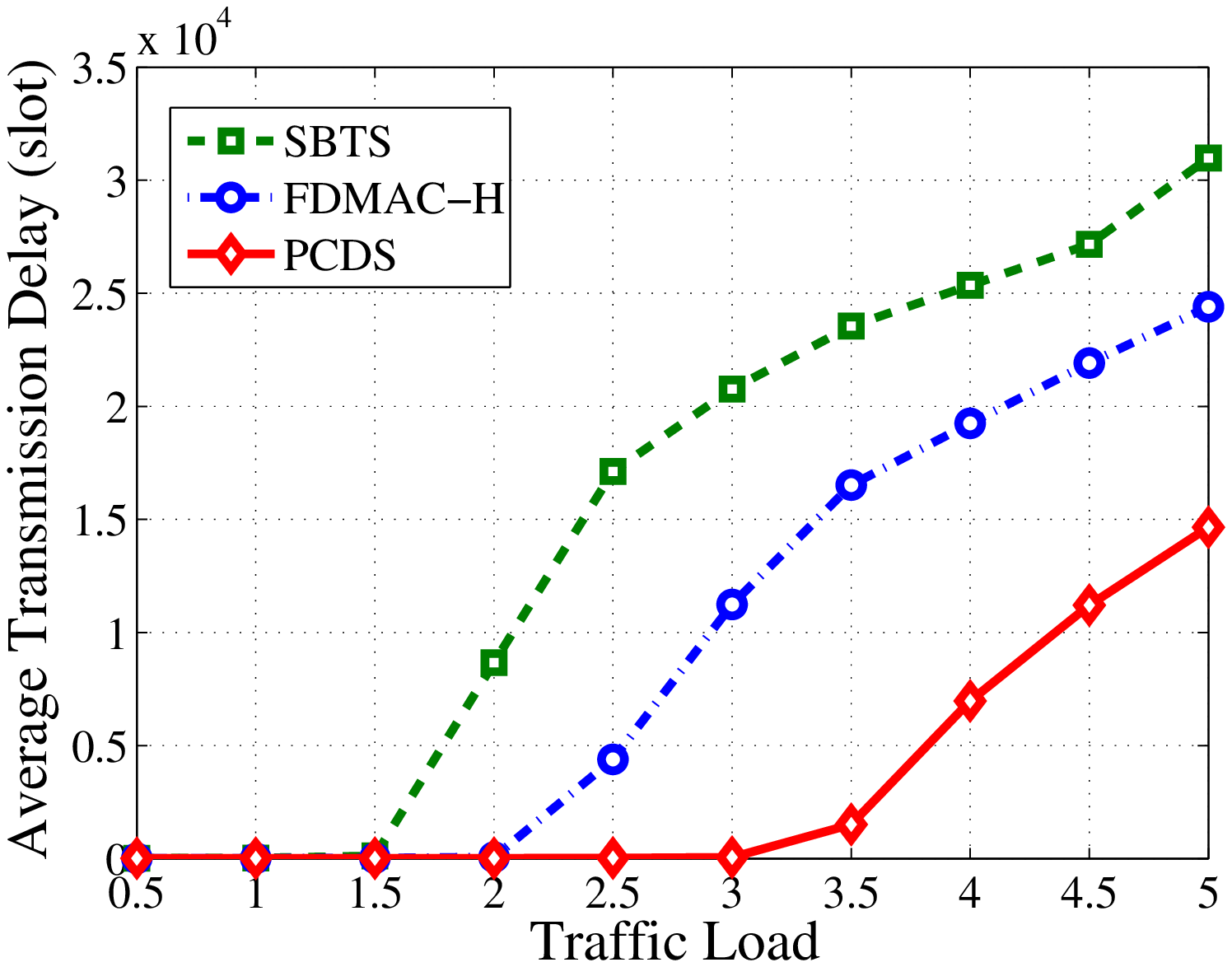}
\centerline{\small (b) IPP traffic}
\end{minipage}%
\caption{Average transmission delay of the three transmission schemes under different traffic loads.}
\label{fig:BD} 
\vspace*{-3mm}
\end{figure}

We plot the average transmission delay of three schemes under different traffic loads in Fig. \ref{fig:BD}. We can observe that the delay curves of the three protocols rise slowly under light traffic loads. The average delay of SBTS and FDMAC-H begin to increase rapidly when the traffic load exceeds 1.5 and 2, while the delay of PCDS doesn't have obvious growth until the traffic load is greater than 3. Under heavy loads, PCDS outperforms SBTS and FDMAC-H significantly. With the traffic loads varying from 3 to 5, PCDS reduces the average delay by about 69.2\% and 68.6\% under Poisson and IPP traffic compared with FDMAC-H, respectively. Compared with SBTS, PCDS reduces the average delay by about 75.5\% and 75.5\% under Poisson and IPP traffic, respectively. In SBTS, since the links from the AP to UEs are adjacent, spatial reuse cannot be exploited. Furthermore, for UEs that have low transmission rates from the AP, their downloading suffers from higher delay. However, by D2D transmissions, PCDS allows UEs to receive packets from other UEs that have received the packets as well as the AP, and the low rate links from the AP to UEs are broken up into multiple hops of high rate, where more efficient spatial reuse can be exploited to improve transmission efficiency and reduce delay. The reason that PCDS outperforms FDMAC-H is that our concurrent transmission scheduling algorithm computes schedules with higher efficiency for the transmission paths selected by Algorithm \ref{alg:path selection}.


\subsubsection{Throughput}

\begin{figure}[htbp]
\begin{minipage}[t]{0.5\linewidth}
\centering
\includegraphics[width=1\columnwidth,height=1.15in]{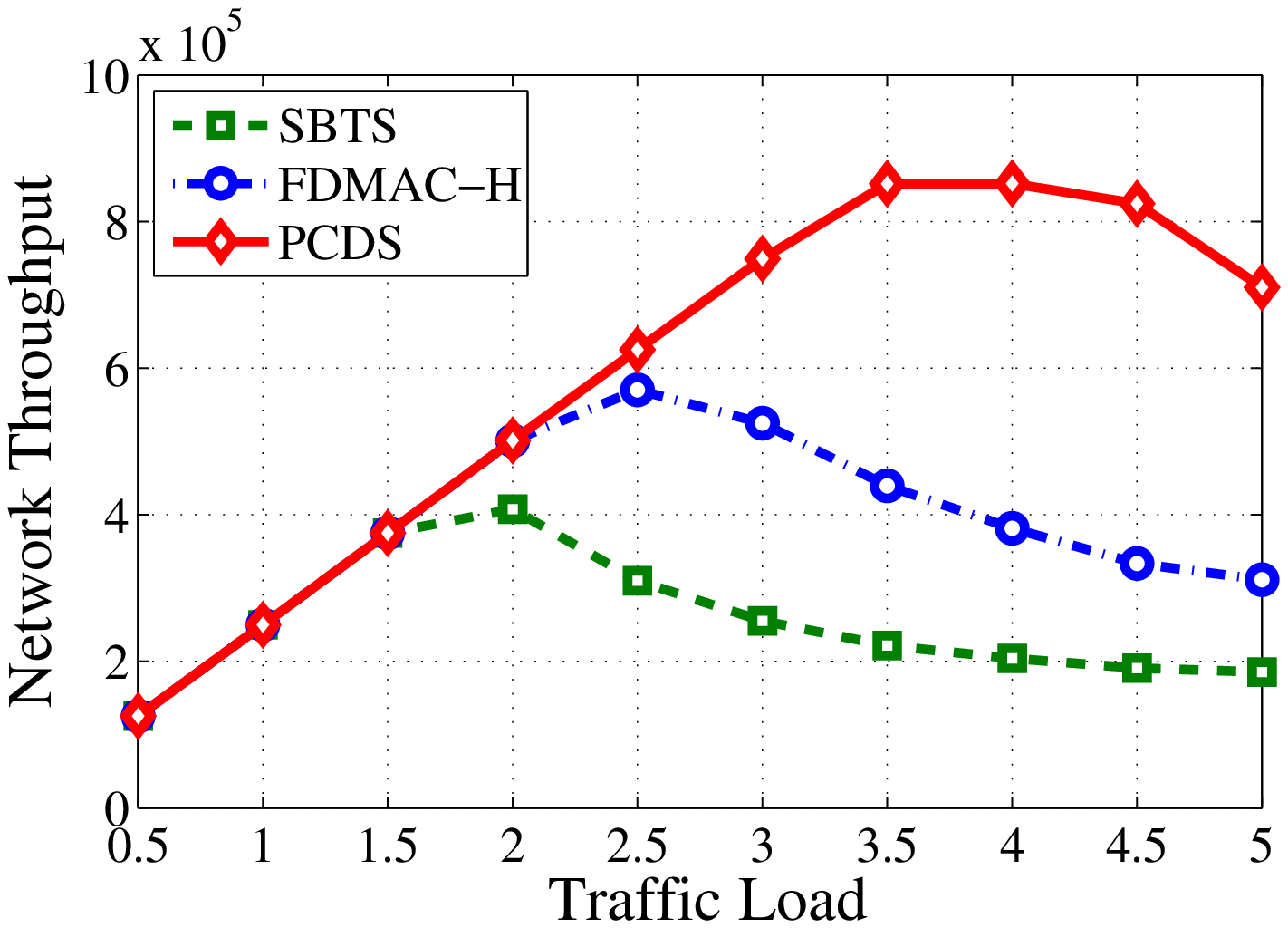}
\centerline{\small (a) Poisson traffic}
\end{minipage}%
\begin{minipage}[t]{0.5\linewidth}
\centering
\includegraphics[width=1\columnwidth,height=1.15in]{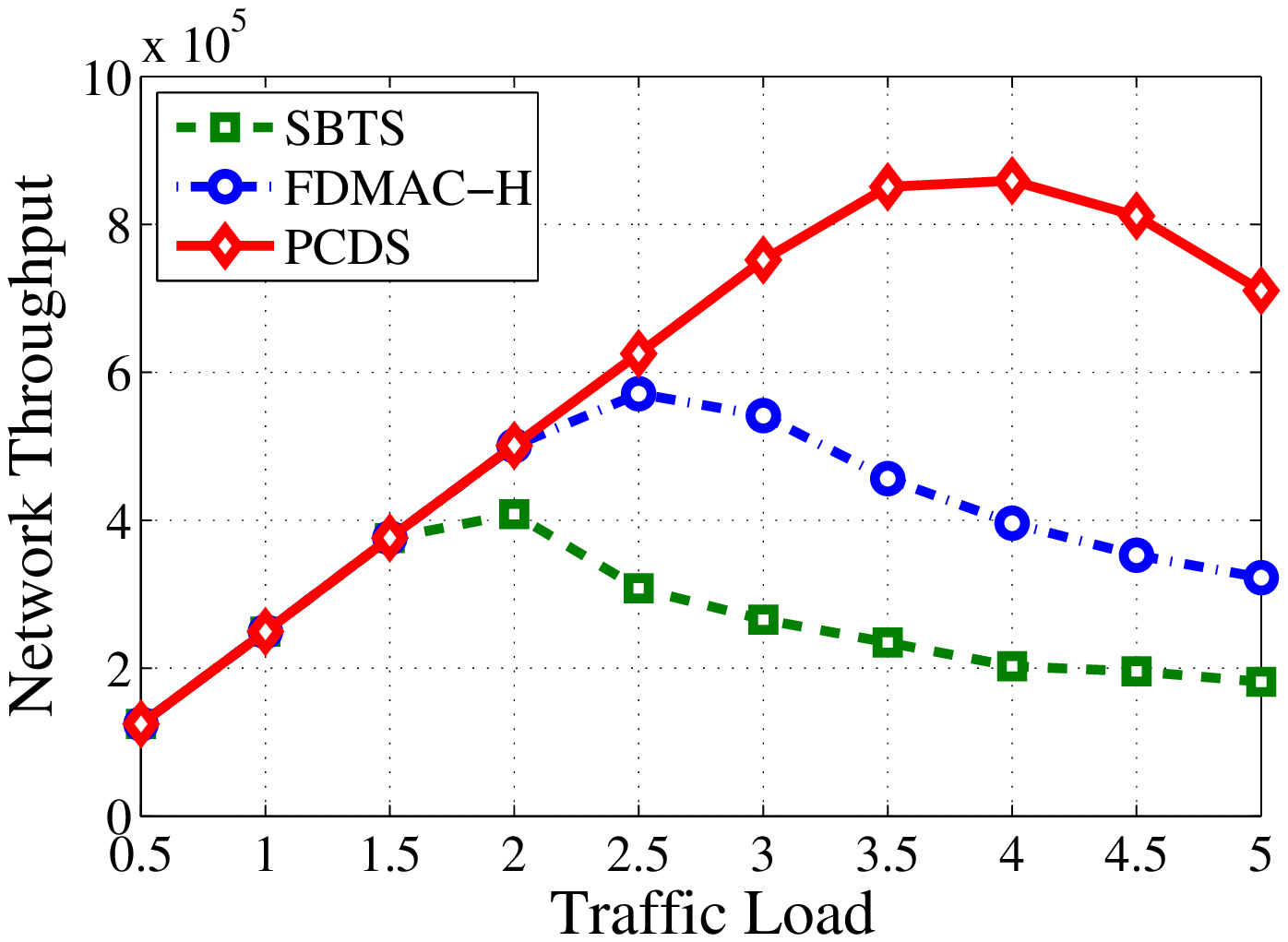}
\centerline{\small (b) IPP traffic}
\end{minipage}%
\caption{Network throughput of the three transmission schemes under different traffic loads.}
\label{fig:NT} 
\vspace*{-3mm}
\end{figure}

The network throughput achieved by the three schemes are plotted in Fig. \ref{fig:NT}. All three protocols have similar performance under light loads, and the throughput increases almost linearly with the traffic load. This is due to the fact that almost all arriving packets are distributed to all UEs successfully within the simulation duration. When the traffic load exceeds 1.5 and 2, the increases of throughput under SBTS and FDMAC-H fail to keep up with the traffic load, and the throughput curves begin to drop when the traffic load is over 2 and 2.5, respectively. However, the curves of PCDS only begin to drop at the traffic load of 4. Compared with SBTS, PCDS increases the network throughput by about 282.5\% on average with the traffic load varying from 3 to 5 under Poisson traffic, and about 275.1\% under IPP traffic. PCDS also increases the network throughput by about 107.2\% on average with the traffic load varying from 3 to 5 under Poisson traffic compared with FDMAC-H, and about 98.5\% under IPP traffic. With the increase of the traffic load, the delay increases, and a considerable number of packets cannot be transmitted within the simulation duration. In addition, there are also more packets that are not counted as successful transmissions due to their delays exceeding the delay threshold. D2D transmissions enable the AP to offload a part of traffic distribution to UEs, and thus the potential of concurrent transmissions is fully unleashed.

\subsubsection{Complexity}

In terms of complexity, the SBTS scheme has much lower complexity compared with PCDS, but its performance is much poorer. FDMAC-H has the same transmission path selection algorithm as PCDS, which is $\mathcal{O}(|\mathbb{U}|^2)$. For the transmission scheduling algorithm, FDMAC-H has the complexity of $\mathcal{O}(|\mathbb{U}|^3)$, which is the same as that of PCDS. Therefore, FDMAC-H has the same complexity as PCDS, but its performance is poorer than PCDS.

\subsection{Impact of the Maximum Number of Hops} \label{S5-3}

In this Section, we investigate the performance of PCDS under different maximum number of hops in the algorithm of transmission
path selection, $H_{max}$, and denote PCDS with $H_{max}$ equal to 2, 3, and 4 by PCDS-2, PCDS-3, and PCDS-4, respectively.

\begin{figure}[htbp]
\begin{minipage}[t]{0.5\linewidth}
\centering
\includegraphics[width=1\columnwidth,height=1.15in]{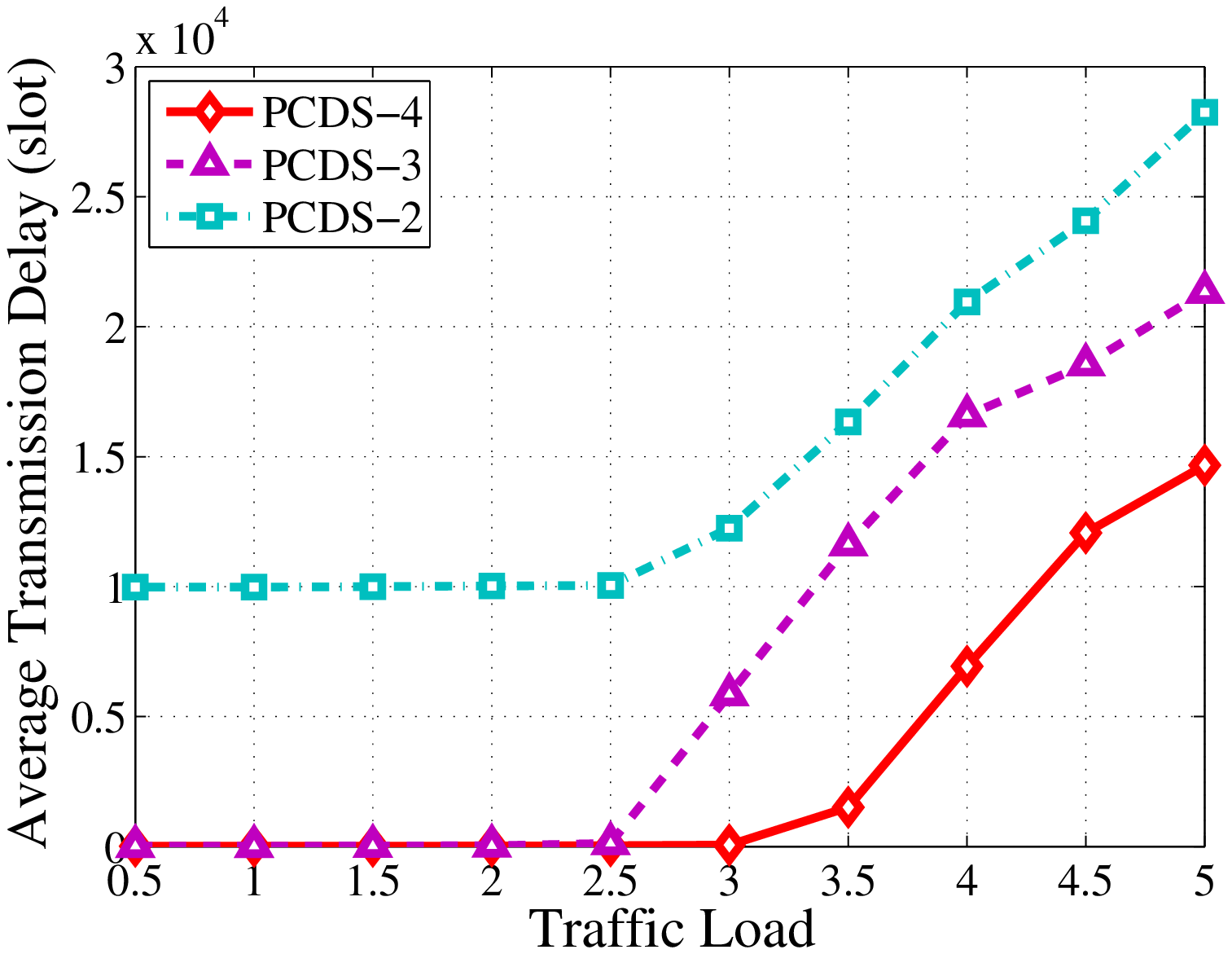}
\centerline{\small (a) Poisson traffic}
\end{minipage}%
\begin{minipage}[t]{0.5\linewidth}
\centering
\includegraphics[width=1\columnwidth,height=1.15in]{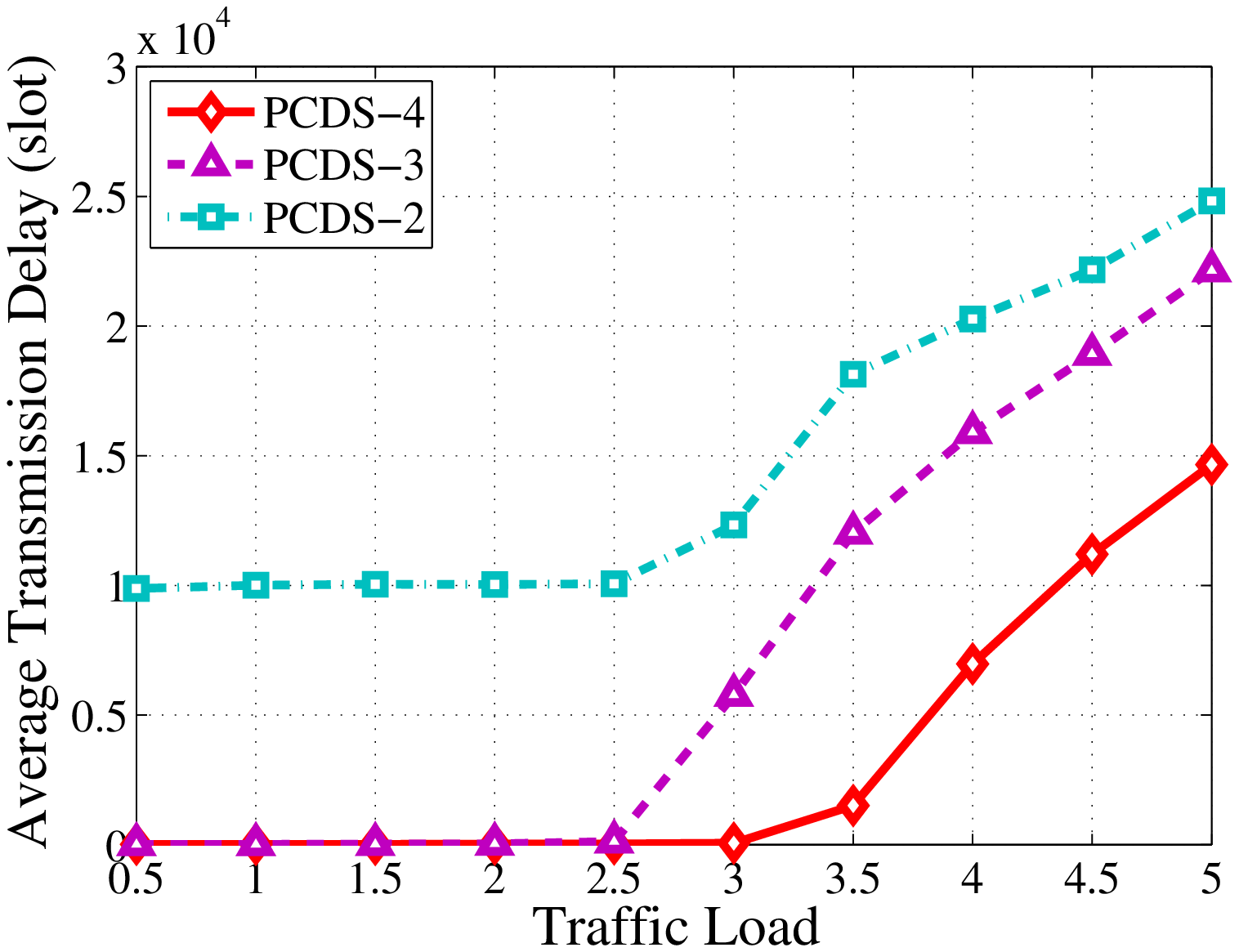}
\centerline{\small (b) IPP traffic}
\end{minipage}%
\caption{Average transmission delay of PCDS with different $H_{max}$.}
\label{fig:BD-H} 
\vspace*{-3mm}
\end{figure}

In Fig. \ref{fig:BD-H}, we plot the average transmission delay of PCDS with different $H_{max}$. Under light traffic loads, PCDS-4 and PCDS-3 have similar performance, but outperform PCDS-2 significantly. Under heavy loads between 3 and 5, PCDS-4 has more significant advantages in terms of delay. PCDS-4 reduces the delay by about 62.1\% on average with the traffic load between 3 and 5 under Poisson traffic compared with PCDS-3, and by about 71\% compared with PCDS-2. With larger $H_{max}$, the links of low channel quality from the AP to UEs are broken up into more hops, and concurrent transmissions can be exploited more fully to improve performance.

\begin{figure}[htbp]
\begin{minipage}[t]{0.5\linewidth}
\centering
\includegraphics[width=1\columnwidth,height=1.15in]{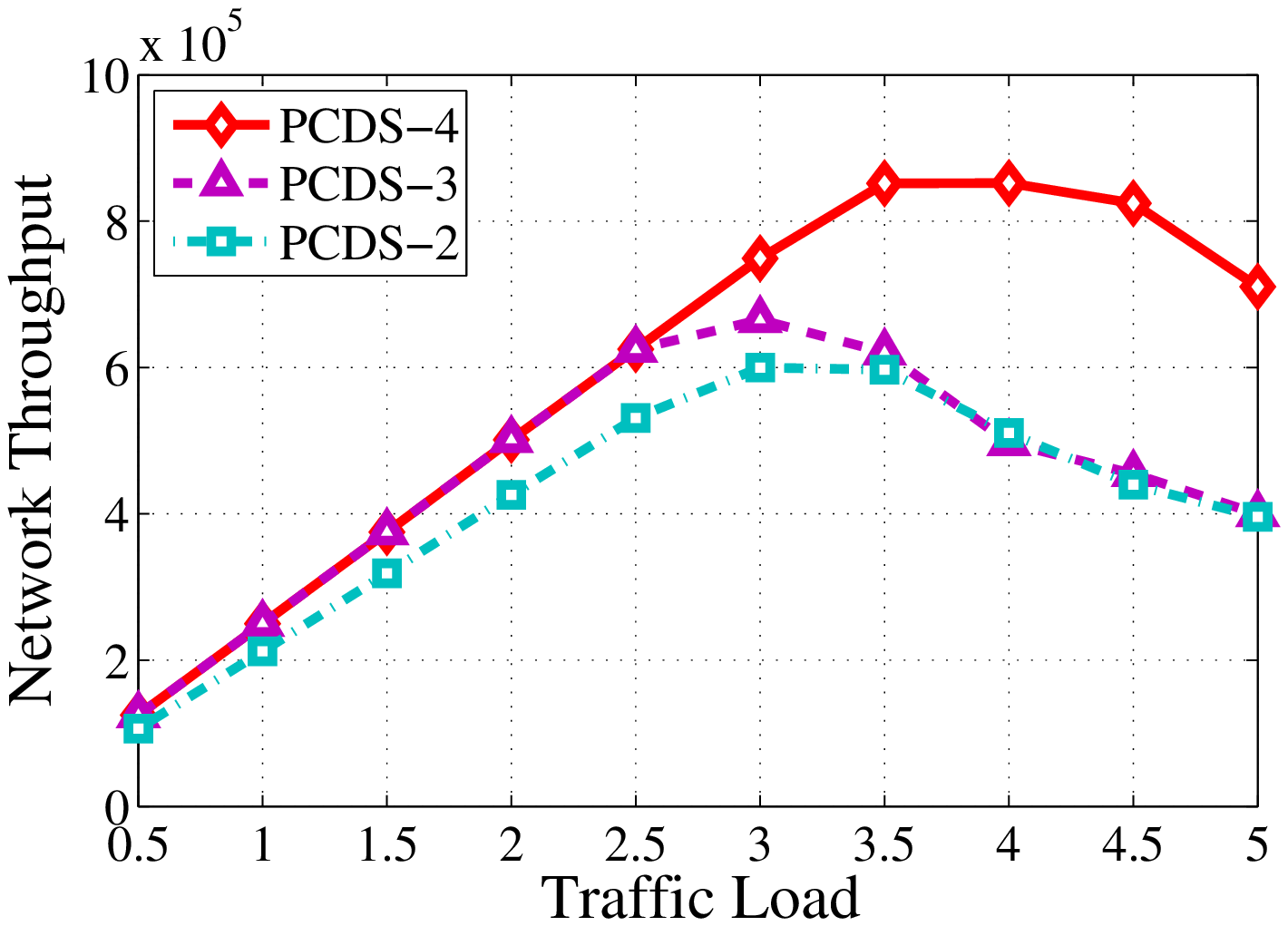}
\centerline{\small (a) Poisson traffic}
\end{minipage}%
\begin{minipage}[t]{0.5\linewidth}
\centering
\includegraphics[width=1\columnwidth,height=1.15in]{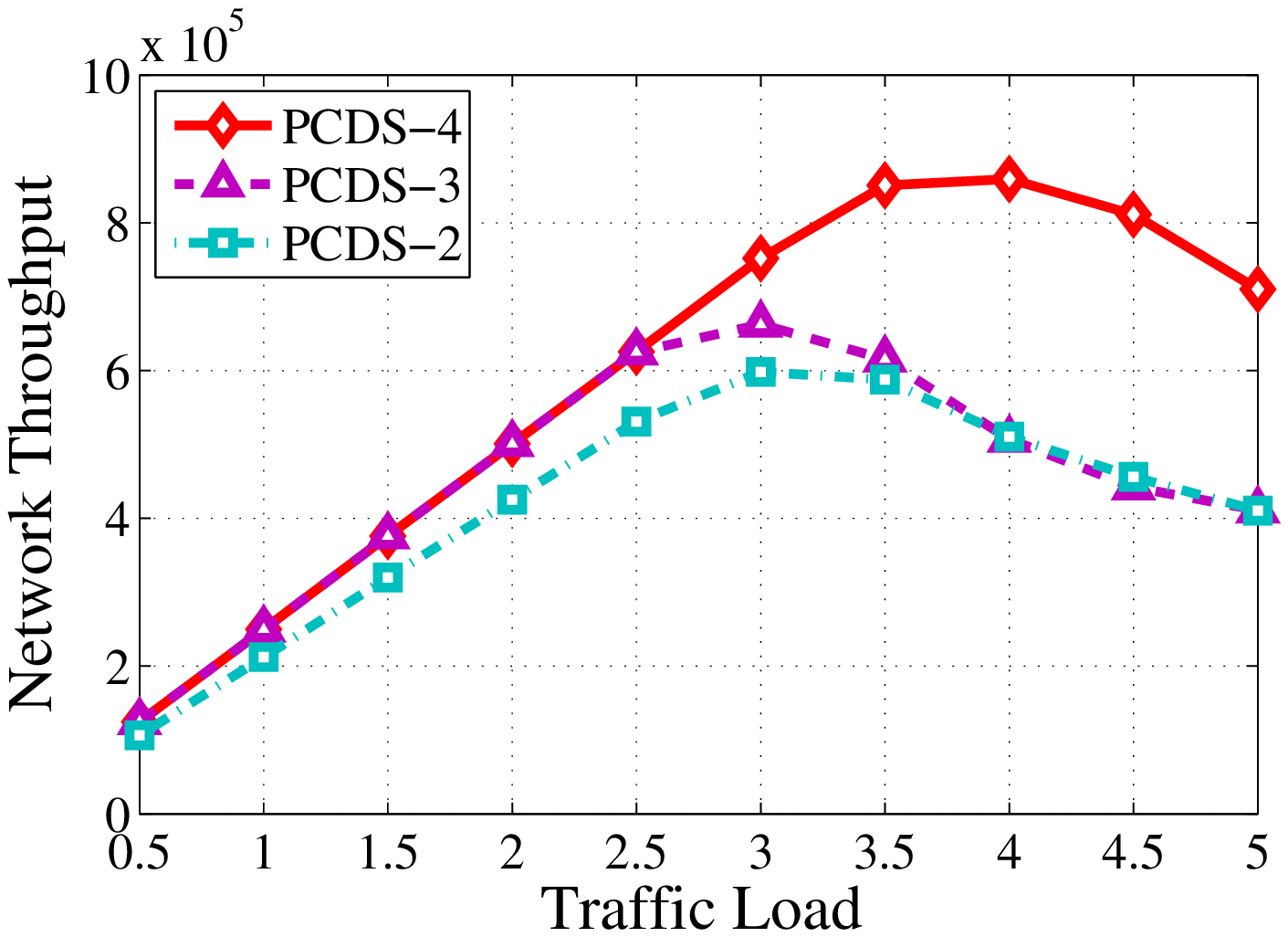}
\centerline{\small (b) IPP traffic}
\end{minipage}%
\caption{Network throughput of PCDS with different $H_{max}$.}
\label{fig:NT-H} 
\vspace*{-3mm}
\end{figure}

We plot the network throughput of PCDS with different $H_{max}$ in Fig. \ref{fig:NT-H}. We can observe that the throughput is consistent with the delay in Fig. \ref{fig:BD-H}. With the traffic load varying from 3 to 5, PCDS-4 increases the network throughput by about 55.5\% on average compared with PCDS-3, and about 57.9\% compared with PCDS-2, respectively, under IPP traffic. With larger $H_{max}$, more traffic downloading is offloaded from the AP to UEs, which exploits D2D transmissions with higher transmission rate to improve network throughput.

\begin{figure}[htbp]
\begin{minipage}[t]{0.5\linewidth}
\centering
\includegraphics[width=1\columnwidth,height=1.15in]{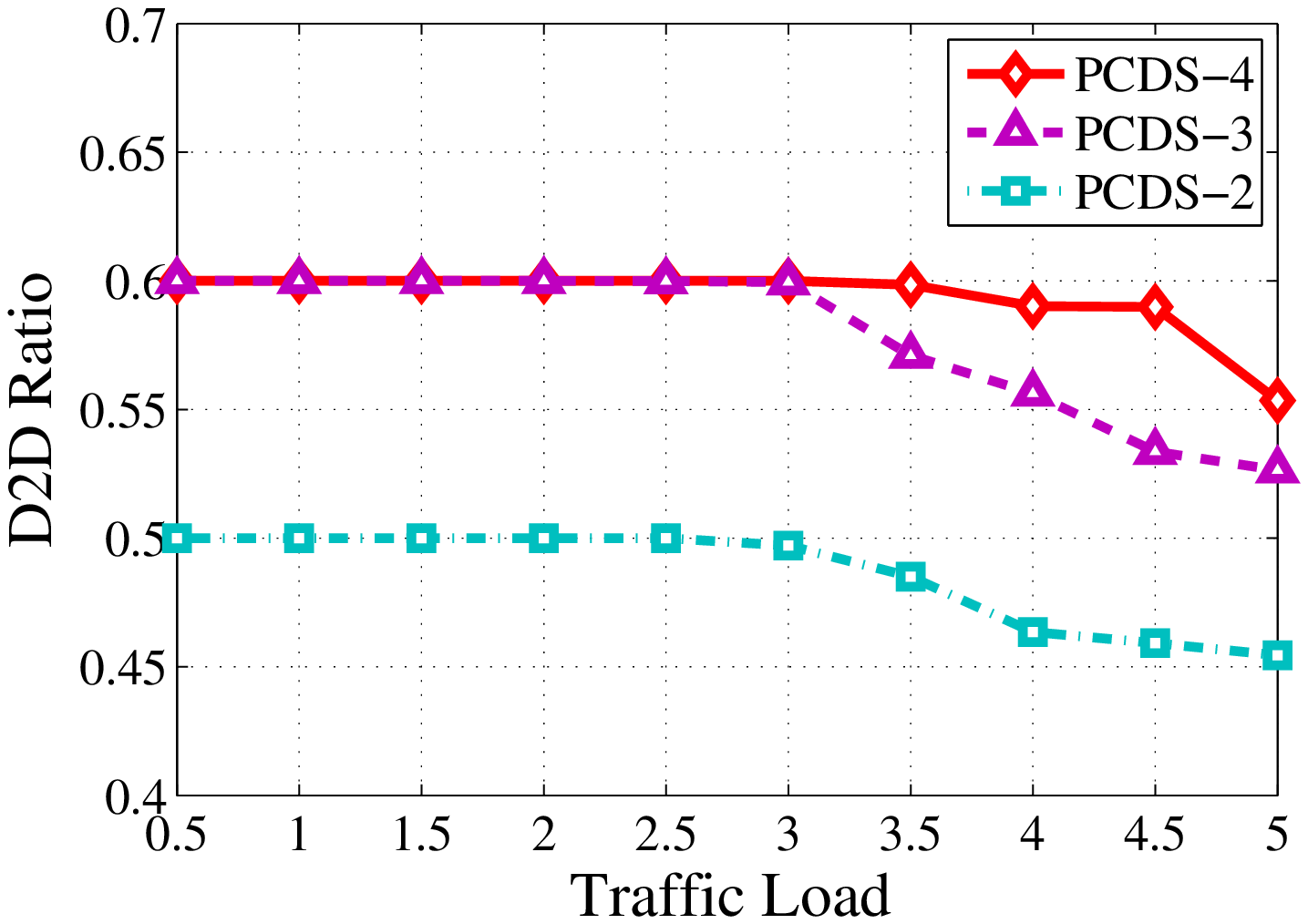}
\centerline{\small (a) Poisson traffic}
\end{minipage}%
\begin{minipage}[t]{0.5\linewidth}
\centering
\includegraphics[width=1\columnwidth,height=1.15in]{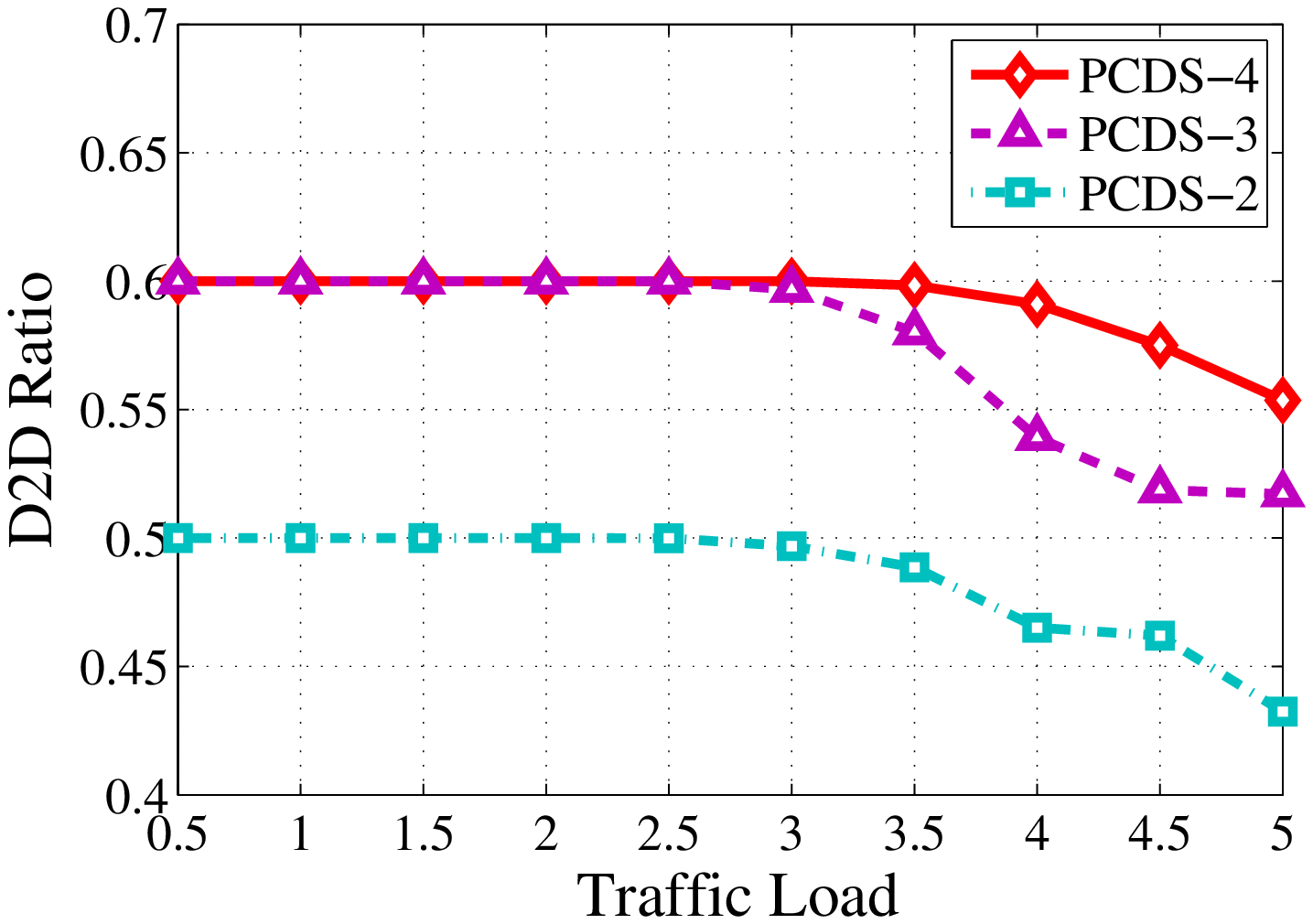}
\centerline{\small (b) IPP traffic}
\end{minipage}%
\caption{D2D Ratios of PCDS with different $H_{max}$.}
\label{fig:R-H} 
\vspace*{-3mm}
\end{figure}

In Fig. \ref{fig:R-H}, we plot the D2D ratios of PCDS with different $H_{max}$. We can observe PCDS-4 and PCDS-3 have higher D2D ratios than PCDS-2, especially under heavy loads. Thus, D2D transmissions are exploited more fully and efficiently in PCDS-4 than PCDS-3 and PCDS-2, which is also consistent with the results in Fig. \ref{fig:BD-H} and Fig. \ref{fig:NT-H}.

Since each UE is only allowed as the downloading source of one neighboring UE once, the number of hops of the selected paths increases as a tolerance of 1 arithmetic progression starting from 1. If the number of UEs can be expanded to the sum of an arithmetic progression, the maximum number of hops of paths can be inferred as ${\frac{{\sqrt {1 + 8|\mathbb{U}|}  - 1}}{2}}$. Otherwise, the maximum possible number of hops of the selected paths is $\left\lceil {\frac{{\sqrt {1 + 8|\mathbb{U}|}  - 1}}{2}} \right\rceil $. Thus, although larger $H_{max}$ usually indicates better performance, the maximum number of hops of transmission paths in Algorithm \ref{alg:path selection} is limited by $\left\lceil {\frac{{\sqrt {1 + 8|\mathbb{U}|}  - 1}}{2}} \right\rceil $. Therefore, $H_{max}$ should be selected according to actual network conditions in practice.

\section{Conclusion}\label{S7} 

In this paper, we proposed PCDS for popular content downloading in mmWave small cells in the 60 GHz band, which exploits both D2D transmission in close proximity and concurrent transmissions to improve transmission efficiency. Transmission path selection is optimized for better use of D2D communications and concurrent transmissions in content downloading. Then a concurrent transmission scheduling algorithm is designed to exploit spatial reuse to improve transmission efficiency. Finally, extensive simulations under various traffic patterns demonstrate PCDS reduces transmission delay and improves network throughput significantly compared with other existing schemes, especially under heavy load. Compared with FDMAC-H, PCDS improves the network throughput by about 102.9\% on average with the traffic load between 3 and 5.




\begin{biography}[{\includegraphics[width=1in,height=1.25in,clip,keepaspectratio]{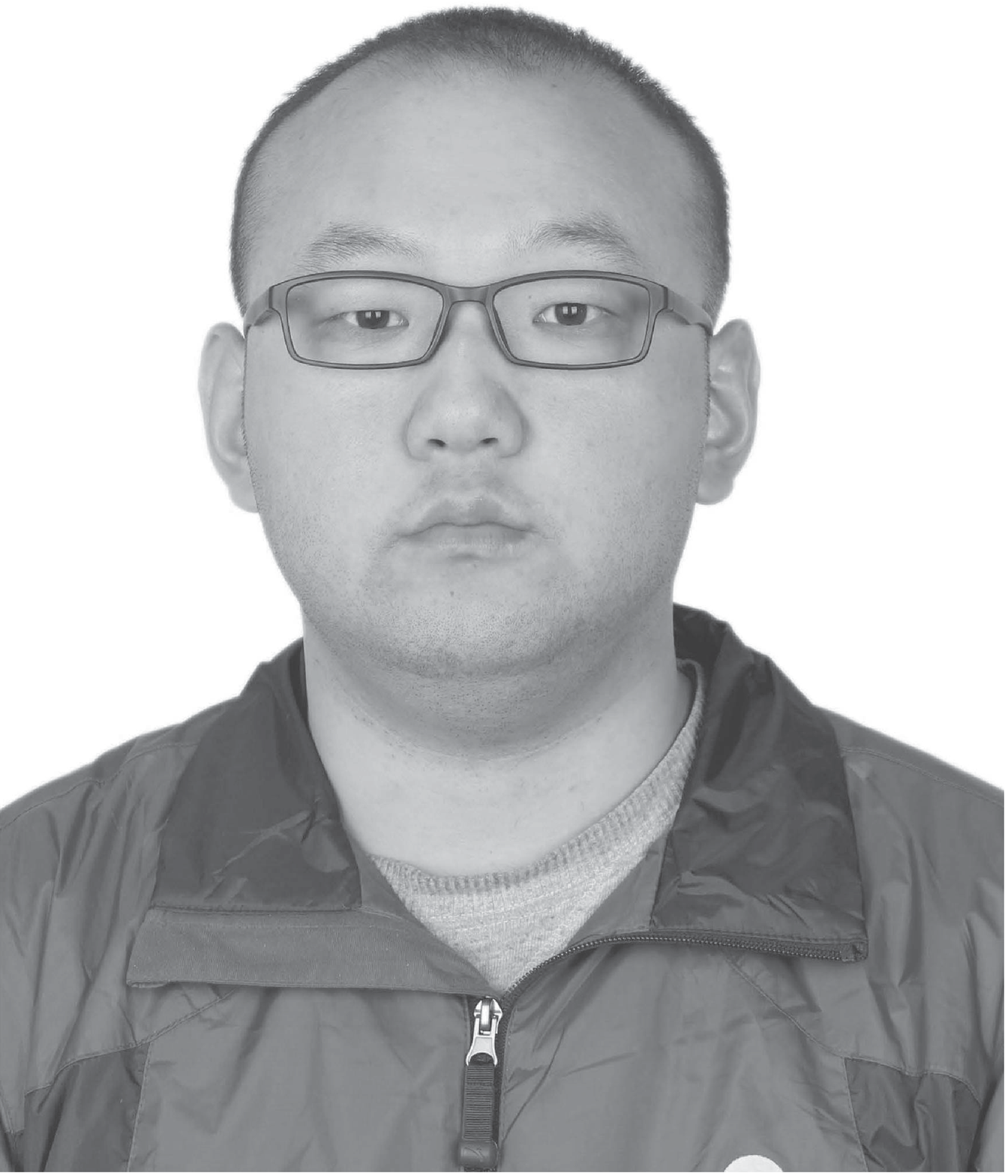}}]{Yong Niu}
received B.E. degree in Beijing Jiaotong University, China in 2011. He is currently working toward
his PhD degree at the Department of Electronic Engineering, Tsinghua University, China. His
research interests include millimeter wave communications, medium access control, and software-defined
networks.
\end{biography}

\begin{biography}[{\includegraphics[width=1in,height=1.25in,clip,keepaspectratio]{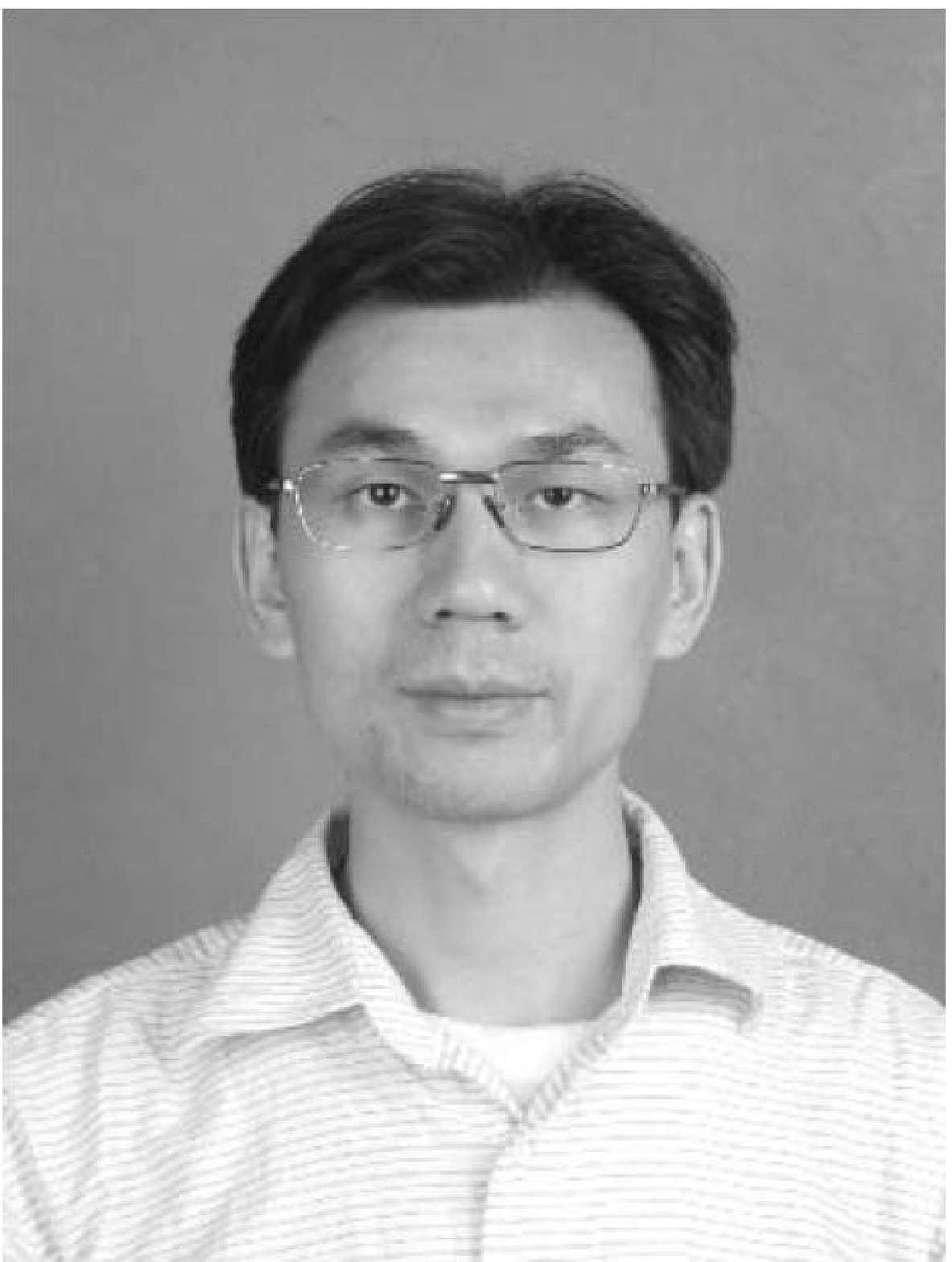}}]{Li Su} received the B.S. degree from Nankai University,
Tianjin, China, in 1999 and Ph.D. degree from Tsinghua University, Beijing, China in 2007
respectively both in electronics engineering. Now he is a research associate with Department of
Electronic Engineering, Tsinghua University. His research interests include telecommunications,
future internet architecture and on-chip network.
\end{biography}

\begin{biography}[{\includegraphics[width=1in,height=1.25in,clip,keepaspectratio]{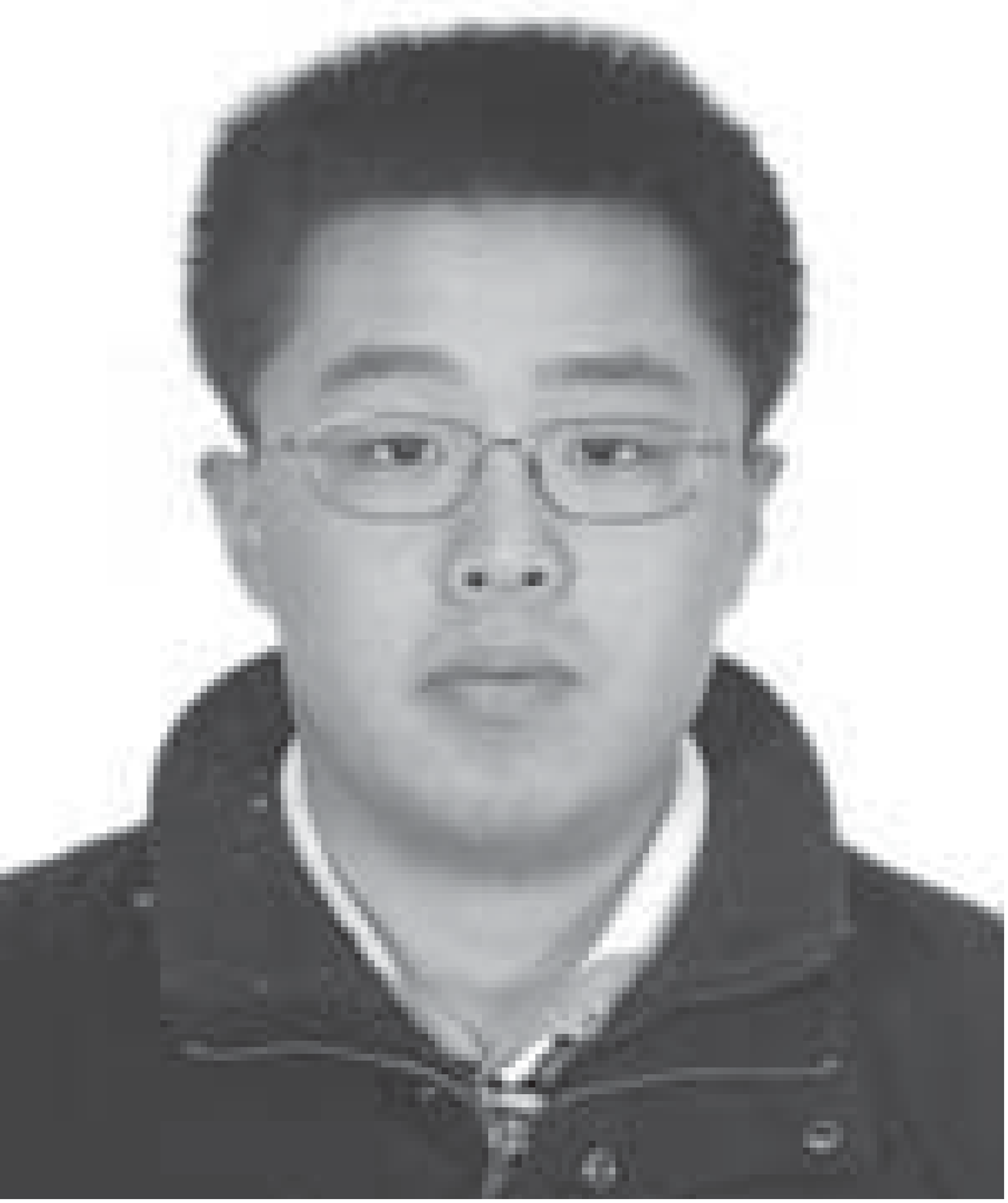}}]{Chuhan Gao}
is an undergraduate student from Tsinghua University, Beijing, China. He is currently working toward the B.E. degree. His research interests include millimeter-wave wireless communications, wireless networks, and software-defined networks.
\end{biography}

\begin{IEEEbiography}[{\includegraphics[width=0.9in,height=1.1in,clip,keepaspectratio]{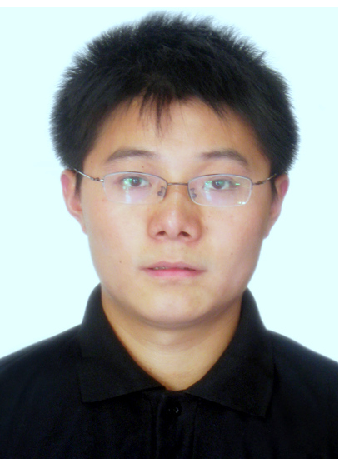}}]{Yong Li}
(M'2009) received his B.S. degree in Electronics and Information Engineering from Huazhong
University of Science and Technology, Wuhan, China, in 2007, and his Ph.D. degree in electronic
engineering from Tsinghua University, Beijing, China, in 2012. During July to August in 2012 and
2013, he worked as a Visiting Research Associate in Telekom Innovation Laboratories (T-labs) and HK
University of Science and Technology, respectively. During December 2013 to March 2014, he visited
University of Miami, FL, USA as a Visiting Scientist. He is currently a faculty member of the
Electronic Engineering at the Tsinghua University.\\ His research interests are in the areas of
networking and communications, including mobile opportunistic networks, device-to-device
communication, software-defined networks, network virtualization, future Internet, etc. He received
Outstanding Postdoctoral Researcher, Outstanding Ph. D Graduates and Outstanding Doctoral thesis of
Tsinghua University, and his research is granted by Young Scientist Fund of Natural Science
Foundation of China, Postdoctoral Special Find of China, and industry companies of Hitachi, ZET,
etc. He has published more than 100 research papers and has 10 granted and pending Chinese and
International patents. His Google Scholar Citation is about 440 with H-index of 11, also with more
than 120 total citations without self-citations in Web-of Science. He has served as Technical
Program Committee (TPC) Chair for WWW workshop of Simplex 2013, served as the TPC of several
international workshops and conferences. He is also a guest-editor for ACM/Springer Mobile Networks
\& Applications, Special Issue on Software-Defined and Virtualized Future Wireless Networks. Now,
he is the Associate Editor of EURASIP journal on wireless communications and networking.
\vspace*{-10mm}
\end{IEEEbiography}

\begin{biography}[{\includegraphics[width=1in,height=1.25in,clip,keepaspectratio]{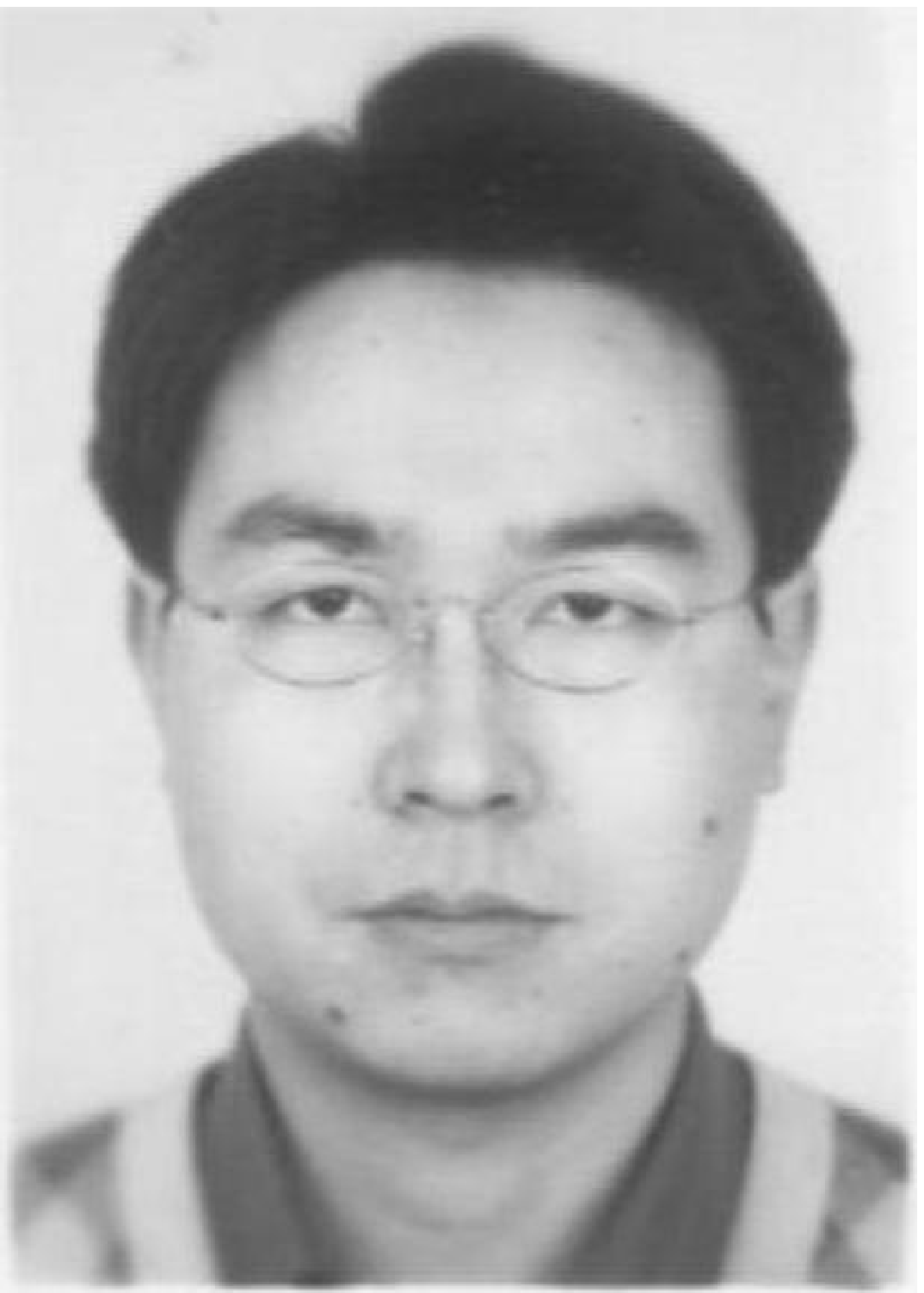}}]{Depeng Jin}
received the B.S. and Ph.D. degrees from Tsinghua University, Beijing,
China, in 1995 and 1999 respectively both in electronics engineering. \\
He is an associate professor at Tsinghua University and vice chair of Department of Electronic
Engineering. Dr. Jin was awarded National Scientific and Technological Innovation Prize (Second
Class) in 2002. His research fields include telecommunications, high-speed networks, ASIC design
and future Internet architecture.
\end{biography}

\begin{IEEEbiography}[{\includegraphics[width=1in,height=1.25in,clip,keepaspectratio]{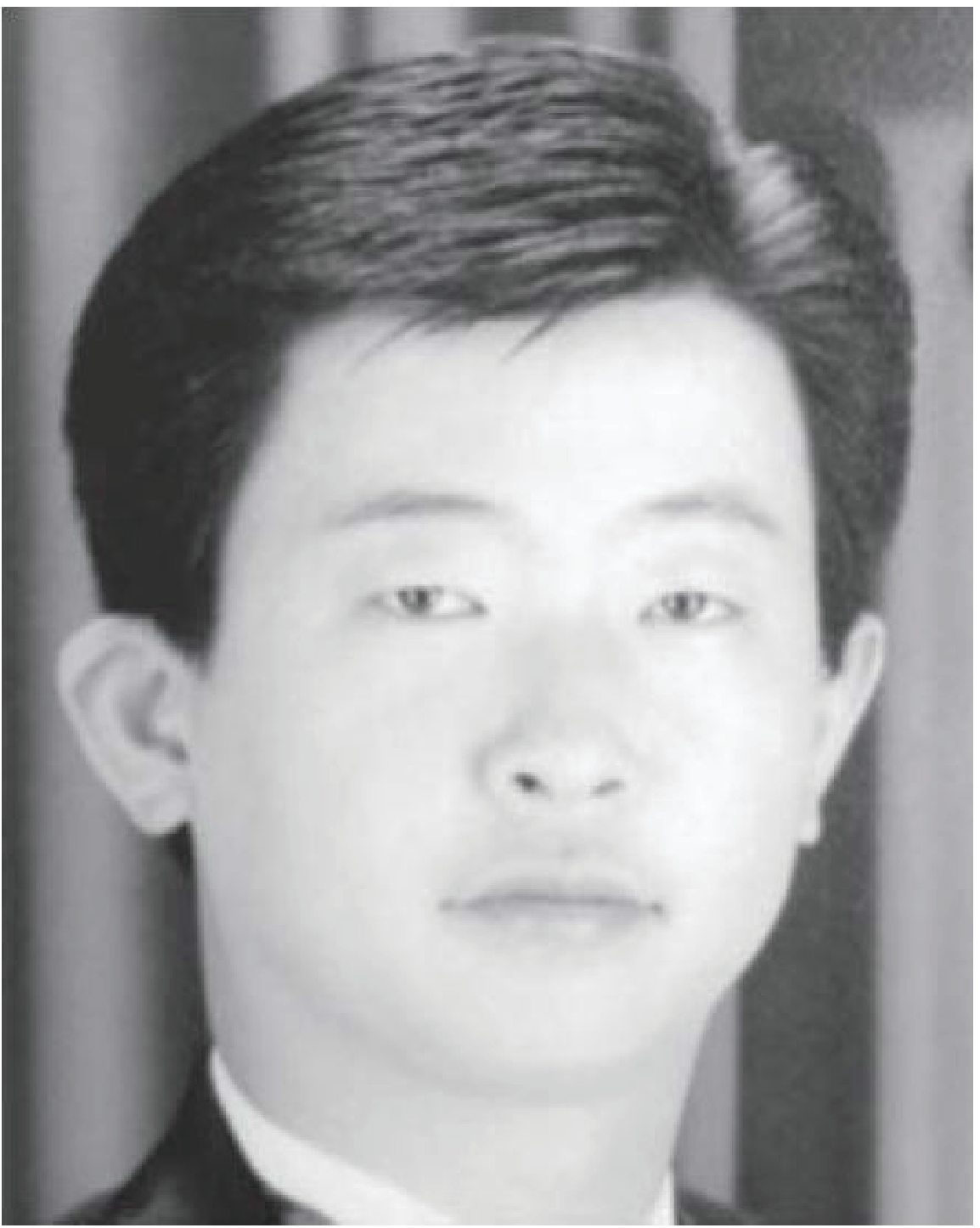}}]{Zhu Han}
(S'01--CM'04--SM'09--F'14) received the
B.S. degree in electronic engineering from Tsinghua
University, Beijing, China, in 1997, and the M.S.
and Ph.D. degrees in electrical engineering from the
University of Maryland, College Park, MD, USA, in
1999 and 2003, respectively.

From 2000 to 2002, he was an R\&D Engineer with
JDSU, Germantown, MD, USA. From 2003 to 2006,
he was a Research Associate with the University of
Maryland. From 2006 to 2008, he was an Assistant
Professor with Boise State University, Boise, ID,
USA. He is currently an Associate Professor with the Department of Electrical
and Computer Engineering, University of Houston, Houston, TX, USA. His research
interests include wireless resource allocation and management, wireless
communications and networking, game theory, wireless multimedia, security,
and smart grid communication. He serves as an Associate Editor for IEEE
Transactions on Wireless Communications since 2010. He received
IEEE Fred W. Ellersick Prize in 2011 and the NSF CAREER award in 2010.

\end{IEEEbiography}

\end{document}